\documentclass[
aps,
pra,
superscriptaddress,
amssymb,
10pt,
reprint,
notitlepage,
nofootinbib,
]{revtex4-1}

\pdfoutput=1
\usepackage[english]{babel}
\usepackage[utf8]{inputenc}

\usepackage{amsmath,amssymb,latexsym}
\usepackage{amsfonts}
\usepackage{braket}

\usepackage{graphicx}

\usepackage[pdfusetitle,colorlinks]{hyperref}

\newcommand{\mode}[1]{\hat{#1}}
\newcommand{\rotmode}[1]{\mode{\tilde{#1}}}
\newcommand{\op}[1]{\mode{#1}}
\newcommand{\ii}{\mathrm i}
\newcommand{\expe}{\mathrm e}
\newcommand{\dif}[1]{\mathrm{d}#1}

\newcommand{\modea}{\mode{a}}
\newcommand{\modeb}{\mode{b}}
\newcommand{\modet}{\mode{c}}
\newcommand{\modebout}{\mode{b}_\text{out}}

\newcommand{\kappaout}{\kappa_\text{out}}
\newcommand{\catket}[2]{\ket{\mathcal{C}^{#1}_{#2}}}
\newcommand{\etaconv}{\eta_\text{conv}}
\newcommand{\etadet}{\eta_\text{det}}

\newcommand{\be}{\begin{equation}}
\newcommand{\ee}{\end{equation}}
\newcommand{\deriv}[1]{\partial_{#1}}
\newcommand{\commute}[2]{\left[#1,#2\right]}

\newcommand{\supplement}{ (Appendix)}

\begin{document}

\title{Schrödinger's catapult: Launching multiphoton quantum states from a microwave cavity memory}
\author{Wolfgang Pfaff}
\email{wolfgang.pfaff@yale.edu}
\author{Christopher J.~Axline}
\author{Luke D.~Burkhart}
\author{Uri Vool}
\author{Philip Reinhold}
\author{Luigi Frunzio}
\author{Liang Jiang}
\author{Michel H.~Devoret}
\author{Robert J.~Schoelkopf}
\affiliation{Departments of Applied Physics and Physics, Yale University, New Haven, CT 06520, USA}
\affiliation{Yale Quantum Institute, Yale University, New Haven, CT 06520, USA}
\date{\today}

\begin{abstract}
Encoding quantum states in complex multiphoton fields can overcome loss during signal transmission in a quantum network. Transmitting quantum information encoded in this way requires that locally stored states can be converted to propagating fields. Here we experimentally show the controlled conversion of multiphoton quantum states, like ``Schr\"odinger cat'' states, from a microwave cavity quantum memory into propagating modes. By parametric conversion using the nonlinearity of a single Josephson junction, we can release the cavity state in $\sim$ 500\,ns, about 3 orders of magnitude faster than its intrinsic lifetime. This `catapult' faithfully converts arbitrary cavity fields to traveling signals with an estimated efficiency of $>90\,\%$, enabling on-demand generation of complex itinerant quantum states. Importantly, the release process can be controlled precisely on fast time scales, allowing us to generate entanglement between the cavity and the traveling mode by partial conversion. Our system can serve as the backbone of a microwave quantum network, paving the way towards error-correctable distribution of quantum information and the transfer of highly non-classical states to hybrid quantum systems.
\end{abstract}

\maketitle

A powerful way to tame complexity when scaling up a quantum system is to construct it as a network. Breaking up the whole into small, testable modules that are connected through well-defined communication channels reduces undesired crosstalk and minimizes the spreading of errors through the system. Therefore, quantum networks have been proposed for quantum information processing (QIP) \cite{Kimble2008} and it has been shown theoretically that there are favorable thresholds for quantum error correction for such modular architectures, even with noisy quantum communication channels \cite{Nickerson2013}. Experiments with multiple platforms are currently underway to realize prototypes of quantum networks \cite{Ritter2012,Monroe2014,Bernien2013}. The key requirement hereby is the ability to interface quantum states stored and processed in network nodes with propagating states that connect the nodes.

Quantum continuous variables (CV) allow versatile and robust encoding of quantum information in higher-dimensional Hilbert spaces. For instance, encoding quantum bits in CV systems can provide the redundancy required to enable quantum error correction \cite{Gottesman2001}. Non-Gaussian CV states that could be used as QIP-enabling resources have been created experimentally in the states of ion motion \cite{Monroe1996} and atomic spins \cite{Signoles2014,McConnell2015}, as well as optical \cite{Ourjoumtsev2006,Neergaard-Nielsen2006} and microwave photons \cite{Deleglise2008,Hofheinz2009,Kirchmair2013,Bretheau2015}. In particular, microwave cavities in superconducting circuits have recently further enabled the storage \cite{Reagor2016} and protection \cite{Ofek2016} of quantum information encoded in non-Gaussian oscillator states. Using these locally stored states as resources in an error-protected, network-based QIP architecture hinges on the ability to interface them with traveling signals (Fig.~1a). However, the controlled mapping of general multiphoton states between a CV quantum memory and traveling signals has so far remained an outstanding challenge.

Here, we experimentally demonstrate the controlled conversion of non-classical multiphoton states from a superconducting microwave cavity to propagating states. Using RF-controlled four-wave mixing in a single Josephson junction we are able to evacuate the cavity about 3 orders of magnitude faster than its natural lifetime. The field is coherently and efficiently upconverted in frequency, and released into a transmission line. This enables on-demand generation of traveling multiphoton quantum states of high fidelity. The excellent temporal control over the conversion process allows shaping of the emitted wavepacket. We use this capability to release only a part of the state stored in the cavity; this partial conversion creates entanglement between the stationary and traveling fields in multiple encodings, confirmed by the observation of non-classical correlations.

\section*{Coupling stationary to propagating microwaves}

To enable communication between network nodes with multiphoton states, a coherent release must meet several important criteria. First, in order to enable distribution of quantum information with high fidelity we require a large ‘on/off ratio’. In the `off'-state the coherence of the memory must be preserved, while the `on'-state allows fast, on-demand release. Further, successful communication requires faithful state mapping, independent of the number of photons. Thus, each photon in the memory should be removed, described by annihilation operator $\modea$, while creating an outgoing photon, described by the creation operator $\modeb^\dagger$. This conversion interaction is effectively a `beam splitter', with Hamiltonian $H_\text{conv} \propto \modea \modeb^\dagger + \text{h.c.}$ Finally, shaping the wave packet is required to enable capture by a receiving node \cite{Cirac1997} and to generate entanglement; we therefore demand precise temporal control over the release process.

\begin{figure}[tp]
    \center
    \includegraphics{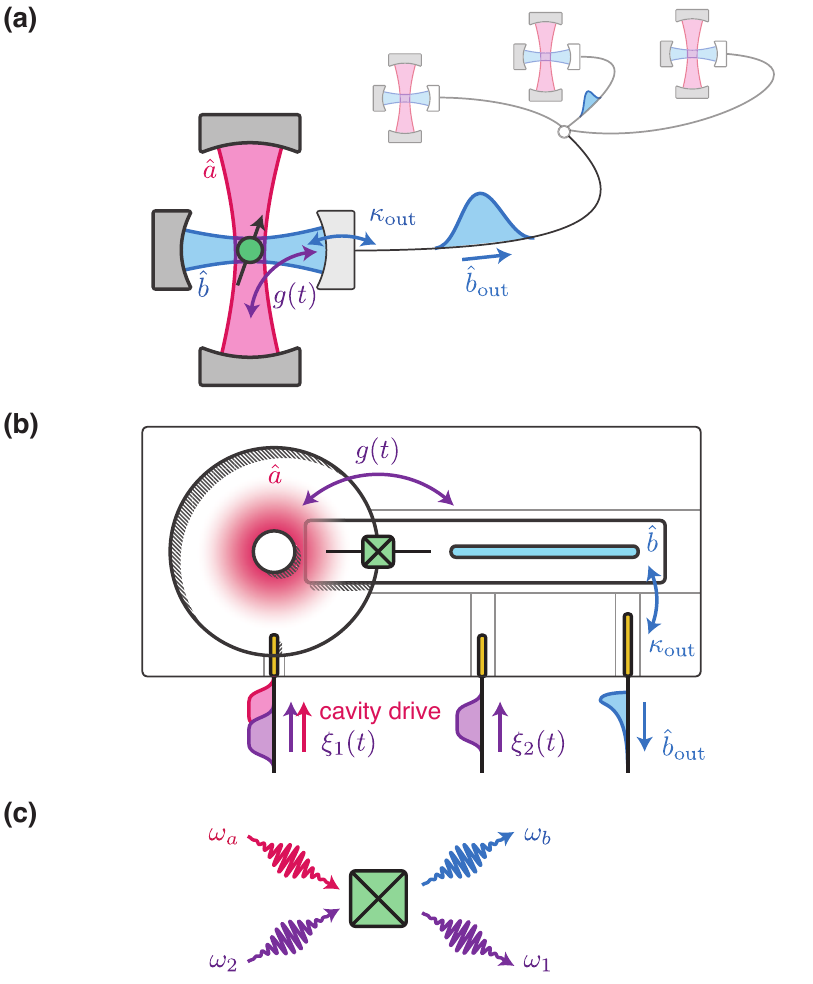}
    \caption{\label{fig:expscheme}
    Nodes in a quantum network using high-Q cavities as memories.
    {\bf a,} We envision an architecture in which each node consists of a high-Q storage mode ($\mode{a}$), an output mode ($\mode{b}$) coupled to a transmission channel ($\mode{b}_\text{out}$) with a rate $\kappa_\text{out}$, and a nonlinear system --- such as an atom (green) --- that allows for a tunable coupling $g(t)$ between $\mode{a}$ and $\mode{b}$. This coupling allows transfer between $\mode{a}$ and $\mode{b}_\text{out}$, and thus transfer of states between nodes.
    {\bf b,} 3D circuit QED implementation of a single node (schematic top view). The storage mode is the fundamental mode of a coaxial cavity (here, $\omega_a/2\pi$ = 4.1\,GHz). The output resonator is a $\lambda/2$ stripline resonator (here, $\omega_b/2\pi$ = 10.0\,GHz), fabricated on the same chip as a single-junction transmon (green) that couples capacitively to both modes. The chip is inserted through a waveguide tunnel. Strongly undercoupled input pins (left and middle) allow application of RF tones (envelopes shown schematically), and signals leave $\mode{b}$ to a transmission line through an output coupler pin (here, $\kappa_\text{out}/2\pi$ = 640\,kHz).
    Qubit control and measurement tones are applied through the input port of the ouput resonator\supplement.
    {\bf c,} Coupling between the resonator modes is realized by tunable pump tones $\xi_{1,2}(t)$ that enable four-wave mixing through the transmon junction. Annihilation and creation of pump photons with frequencies $\omega_{1,2}$ result in upconversion from $\mode{a}$ to $\mode{b}$ if $|\omega_1 - \omega_2| = |\omega_a - \omega_b|$.
    }
\end{figure}

Several approaches have been used to map stationary onto traveling states in superconducting quantum circuits. Tuning the coupling between a superconducting artificial atom and an output mode allows the generation and shaping of single photons \cite{Houck2007,Eichler2012,Srinivasan2014,Pechal2014}. Dedicated coupler elements such as flux-tunable couplers \cite{Yin2013,Pierre2014} or parametric converters \cite{Flurin2015} have been used to release single photons and `quasi-classical' Gaussian CV states. However, no such interface for non-Gaussian CV quantum states has been shown experimentally as of yet.

Our strategy for realizing an interface between arbitrary stationary and traveling multiphoton quantum states is sketched in Fig.~1a. We aim to couple a storage cavity mode, $\modea$, and an output mode, $\modeb$, by a nonlinear element that enables photon conversion between them. This conversion could, for instance, be achieved with parametric amplifiers and converters \cite{Flurin2015}; however, direct coupling of such an element to a long-lived quantum memory has not yet been shown. We instead couple the modes using a transmon artificial atom in the strongly dispersive regime of cavity QED \cite{Schuster2007}; the single Josephson junction of the transmon provides the required nonlinearity, while preserving cavity coherence on the order of milliseconds \cite{Reagor2016}. Our experimental scheme is shown in Fig.~1b. A high-Q superconducting cavity and a low-Q output resonator that are strongly detuned ($|\omega_a - \omega_b|/2\pi \approx 6\,\text{GHz}$) are both coupled to the transmon. The output mode is further coupled to a transmission line mode, $\modebout$, with rate $\kappaout = 1/250$\,ns, where the emitted signals are amplified and recorded\supplement. This configuration enables a long memory life time, $\kappa_0 = 1/450\,\mu\text{s}$, while still allowing for fast readout and control of arbitrary quantum states \cite{Heeres2016}.

Crucially, the nonlinearity induced by the transmon allows conversion of multiphoton states between the memory and output with a large on/off ratio. The Hamiltonian describing the coupling between the modes is given by \cite{Nigg2012}
\begin{equation}
    \label{eq:cosine-hamiltonian}
    H/\hbar = -E_\text{J} \cos(\phi_a(\modea + \modea^\dagger) +
        \phi_b(\modeb + \modeb^\dagger) +
        \phi_c(\modet + \modet^\dagger)),
\end{equation}
where $\modet$ is the annihilation operator for the transmon mode, $E_\text{J}$ is the Josephson energy, and $\phi_i$ is the zero-point fluctuation of flux associated with the respective mode. Because the cosine-coupling enables all four-wave mixing processes that conserve energy, we can create interactions between the strongly detuned resonator modes by applying pump tones. In particular, two pumps with a frequency difference matching the detuning of the resonators enable the conversion Hamiltonian
\begin{equation}
    \label{eq:H_conv}
    H_\text{conv}/\hbar = g(t)\modea \modeb^\dagger + g^*(t)\modea^\dagger \modeb,
\end{equation}
where the coupling $g(t) = E_\text{J}\phi_a^2\phi_b^2 \xi^*_1(t) \xi_2(t)$ can be controlled by the pump strengths $\xi_{1,2}(t)$ (Fig.~1c). Note that the dressed transmon mode $\modet$ does not directly participate in this conversion. From the output mode, photons converted from $\modea$ to $\modeb$ leak into the transmission line. For $g \ll \kappaout$, this results in an effective damping of $\modea$ with rate $\kappa = 4g^2/\kappaout$; the fastest achievable damping is given by the bandwidth of the output mode, $\kappaout$, corresponding to a maximum on/off ratio of the decay that exceeds $10^3$\supplement.

\section*{Cavity evacuation}

We first explore the maximum damping rate we can induce with pump tones of varying strength (Fig.~2a). We prepare the Fock state $\ket{1}$ in the cavity, and then monitor the cavity population over time while applying the pumps. The pump frequencies are tuned on resonance with the conversion process of Eqn.~\ref{eq:H_conv} using $\omega_a - \omega_1 = \omega_b - \omega_2 = 2\pi \times (30-50)\,\text{MHz}$. Increasing the pump strength allows us to tune the cavity energy decay rate from its intrinsic value of $\kappa_0 = 1/0.45\,\text{ms}$ to $\kappa \approx 1/0.5\,\mu\text{s}$ for $g/2\pi = 207\,\text{kHz}$, the maximum conversion rate achievable with the available pump power (Fig.~2b). At this point, $g \approx 0.3 \times \kappaout$, and the decay becomes limited by the bandwidth of the output mode. We can thus use this ``Q-switch'' to evacuate the storage mode with an on/off ratio approaching $10^3$.

\begin{figure}[tp]
    \center
    \includegraphics{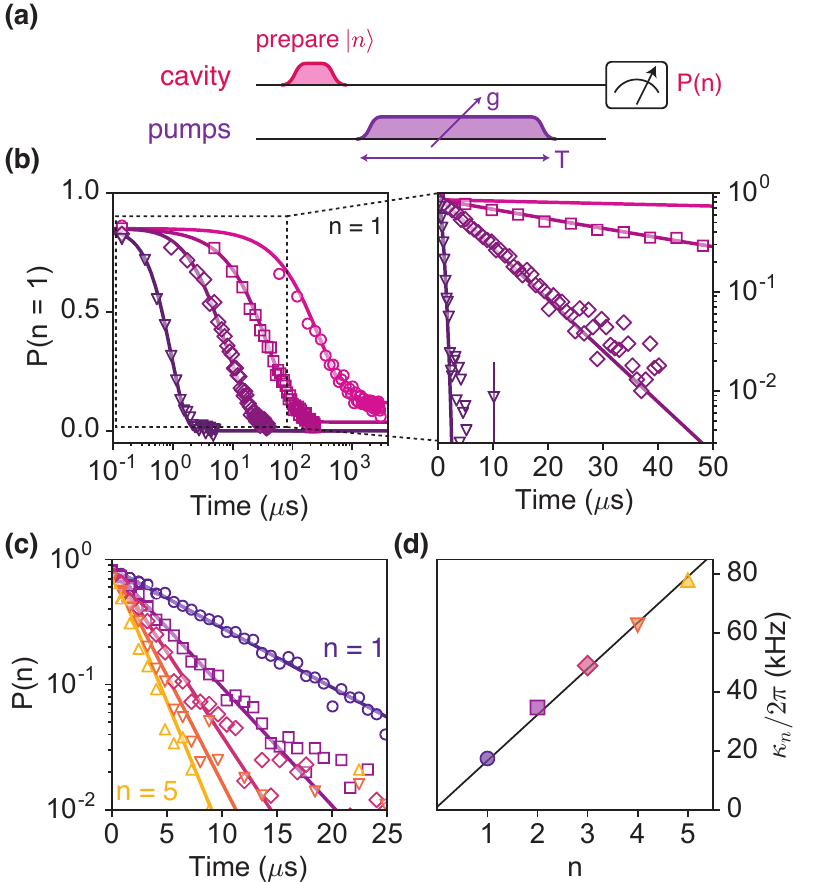}
    \caption{\label{fig:qswitch}
    Cavity damping by mode-conversion.
    {\bf a,} Measurement scheme. After preparing the cavity in a Fock state, we monitor the population as a function of time for different pump strengths. The pump tones have constant amplitude in time, up to a smooth ring-up and ring-down.
    {\bf b,} Decay of the single-photon state $\ket{1}$. $g/2\pi$ = 0 (circles), 25\,kHz (squares), 54\,kHz (diamonds), and 207\,kHz (triangles). Solid lines: for $g$ = 0, exponential fit, yielding the natural decay time; for $g > 0$, theoretical prediction based on independently calibrated pump parameters. For large $g$ the decay is not simply described by a single exponential\supplement. The last datapoint for the fastest decay shows the average and standard deviation for the residual cavity population, consistent with the vacuum state ($P(1) = 0.01 \pm 0.01$).
    {\bf c,} Decay of number states $\ket{n}$, with $n=1..5$; $g/2\pi$ = 54\,kHz. Solid lines are single-exponential fits $P(n) \propto \exp(-\kappa_n t)$.
    {\bf d,} Extracted decay rates $\kappa_n$. Solid line is a linear fit to $\kappa_\text{loss} + n \kappa$, where $\kappa_\text{loss}$ is the independently measured loss rate\supplement.
    }
\end{figure}

This Q-switch is very close to an ideal damping of the memory. It cools the cavity close to the vacuum, with a residual population of $\bar{n} \lesssim 0.01$, the noise floor of our measurement. We can therefore use the conversion as a fast reset, which is a useful tool for experiments with long-lived quantum memories \cite{Wang2016,Heeres2016}. Further, the measured decay of the cavity population is in excellent agreement with predictions based on theory and independent calibrations of the pump strengths. This agreement, together with the absence of any significant heating in the system, suggests a very high conversion efficiency from the storage to the output mode. We estimate that the loss rate into undesired channels, $\kappa_\text{loss}$, is about $0.01 \times \kappa$, corresponding to an expected inefficiency of the conversion of $1-\eta_\text{conv} \approx 0.01$\supplement.

We verify that the cavity evacuation is independent of the input state. We prepare larger Fock states and monitor the population of the input Fock state, $P(n)$ (Fig.~2c). For state-independent damping of a harmonic oscillator, with only a single-photon decay operator $\modea$, we expect the state $\ket{n}$ to decay with a rate $\kappa_n = n \kappa$. From exponential fits to the decay of $P(n)$ we find very good agreement with this linear behavior (Fig.~2d). For larger $n$ we expect that $\kappa_n$ will gradually decrease due to the Kerr effect \cite{Kirchmair2013}. For $n \leq 5$ we find a deviation of $\kappa_n$ from $n \kappa_1$ of $\leq$ 6\%, and therefore state independence is a good approximation\supplement. This can be improved further by reducing the magnitude of the Kerr effect through adjustment of sample parameters.

\section*{Traveling multiphoton quantum states}

\begin{figure*}[tp]
    \center
    \includegraphics{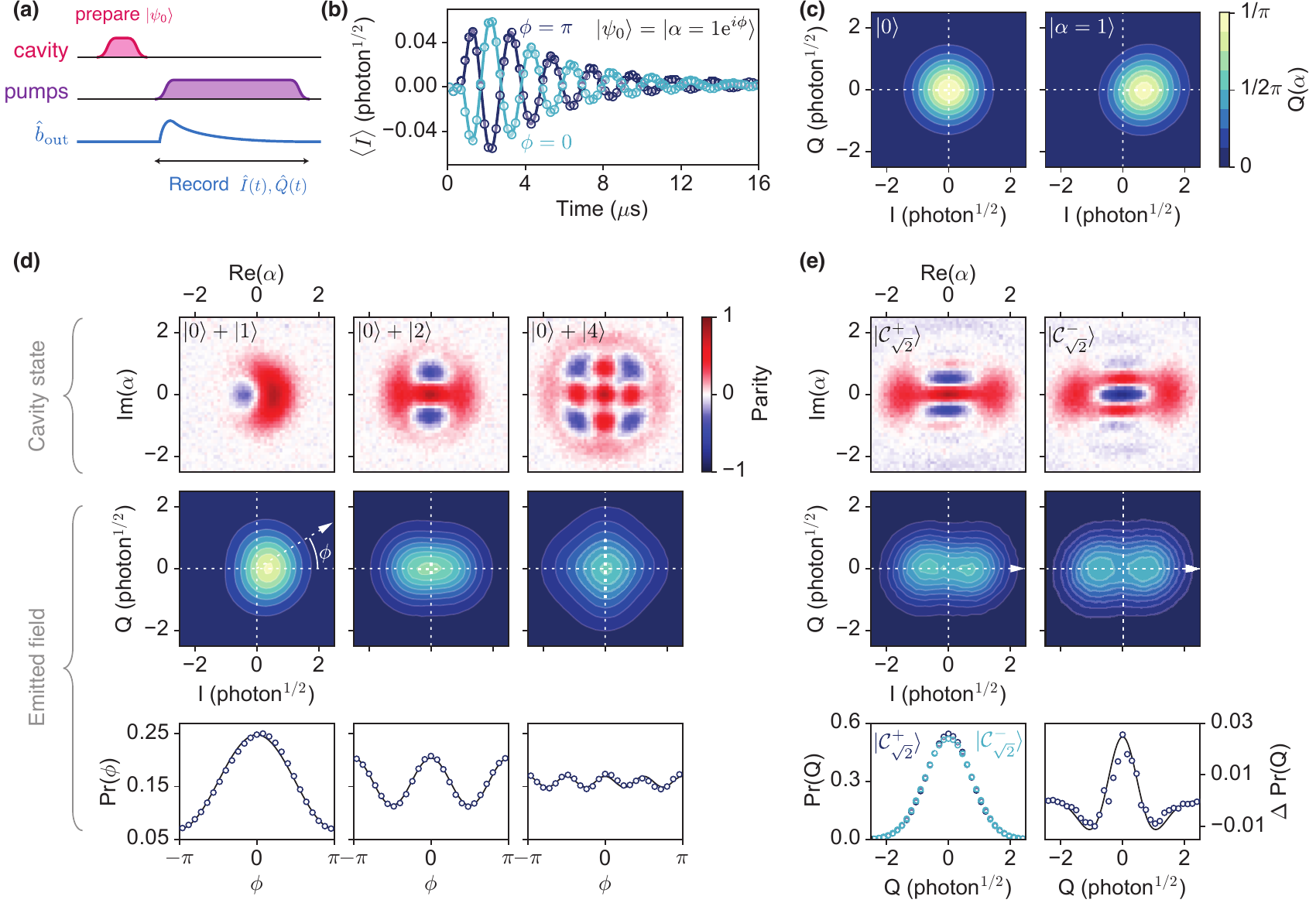}
    \caption{\label{fig:coherence}
    Traveling multiphoton quantum states.
    {\bf a,} After preparing a cavity state, we monitor the output field in heterodyne detection while applying the pump tones.
    {\bf b,} Averaged in-phase signal for two coherent states with $\bar{n}=1$ and opposite phases. $g/2\pi$ = 125\,kHz for this data. Solid lines: fit to a sum of two exponentials, $\langle I(t)\rangle \propto \exp(-\kappa t) - \exp(-\kappa_\text{out} t)$.
    {\bf c,} Q-functions for vacuum and a coherent state obtained by integrating quadrature data in time and computing normalized histograms.
    {\bf d,} Fock state superpositions $(\ket{0}+\ket{n})/\sqrt{2}$. {Top panel:} Measured Wigner function of the prepared cavity state (not corrected for imperfect readout). {Middle:} Q-function of the traveling signal, not corrected for detection loss. {Bottom:} radially integrated Q-function, $\mathrm{Pr}(\phi) = \int r Q(r,\phi) \dif{r}$. Solid line: expected contrast for the ideal state, taking into account the detection efficiency.
    {\bf e,} Even and odd cat states, $\ket{\mathcal{C}^{\pm}_{\sqrt{2}}}$.
    {Lower left:} marginals $\mathrm{Pr}(Q)$, obtained by integrating over $I$ (arrows indicate direction of integration). {Lower right:} Difference between the marginals (odd subtracted from even). Solid line: ideal case.
    All Q-function data have been taken with $g/2\pi$ = 164\,kHz. Fock state superposition Q-functions were taken with $10^7$ samples; Cat state Q-functions with $10^6$ samples.
    }
\end{figure*}

To determine whether cavity states are mapped faithfully onto traveling signals we characterize the field emitted during conversion. We prepare a cavity state and record the field using heterodyne detection with a quantum-limited amplifier \cite{Bergeal2010} (Fig.~3a). The averaged in-phase signal, $\langle I(t) \rangle$, from releasing a coherent state with average photon number $\bar{n}=1$ is shown in Fig.~3b. Because the output mode has a finite bandwidth, we observe an exponential rise of the signal at rate $\kappaout$, followed by an exponential decay with the induced decay rate $\kappa$. The emitted signal clearly retains coherence with the cavity state, made visible as an oscillation by demodulating with a small detuning from the output frequency. Importantly, the amplitude of the oscillations is consistent with a high conversion efficiency from the cavity to the output mode. By calibrating the signal amplitude in terms of the number of photons emitted by the output resonator, we estimate that the propagating field contains $1 \pm 0.15$ photons; this is in agreement with our expectation of a small inefficiency in the conversion\supplement.

A crucial requirement for our interface is that non-classical states are preserved faithfully in the conversion process. Because the averaged signal vanishes for most states of interest, we compute a probability distribution in phase space. By integrating each measurement record $I(t) + \ii Q(t)$ in time and histogramming the results, we directly obtain the Husimi Q-function \cite{Leonhardt1997,Eichler2012}. The measured Q-functions for the vacuum and a coherent state with $\bar{n}=1$  are shown in Fig.~3c. Finite loss in our detection circuitry leads to a smoothing and shrinking of the distribution, the extent of which is determined by the detection efficiency $\eta_\text{det}$ \cite{Leonhardt1997}. The coherent state is thus not centered at $|\alpha=1|$, but is moved closer to the origin. Having established a conversion efficiency of close to unity, the position of the coherent state population in the Q-function is a direct measure of our detection efficiency; assuming no loss in the conversion we obtain $\eta_\text{det} = 0.43 \pm 0.04$, consistent with the expected performance of a JPC\supplement. Although non-classical features are thus blurred by the detector, state-essential signatures are preserved in the raw data; knowledge of the detection efficiency allows us to quantitatively confirm the faithful release of quantum states. We illustrate this using two classes of non-Gaussian oscillator states.

The first class are Fock state superpositions of the form $(\ket{0}+\ket{n})/\sqrt{2}$, which display an $n$-fold symmetry in their quasiprobability distributions. For a set of such states we show the Wigner function of the cavity state, measured directly after preparation \cite{Bertet2002}, and the Q-function of the released field; from comparison it is clear that the two distributions share the same symmetry. For additional clarity, we integrate the Q-function radially to obtain a probability distribution as a function of angle, $\text{Pr}(\phi)$. In this representation it can be seen that the symmetry is fully preserved; the contrast is as expected, given our detection efficiency.

A second class of states of particular interest for CV quantum information processing are ``Schr\"odinger cat'' states of the form $\catket{\pm}{\alpha} = \mathcal{N} (\ket{\alpha} \pm \ket{-\alpha})$, which are eigenstates of photon number parity. We create and release the even ($+$) and odd ($-$) parity coherent-state superpositions $\catket{\pm}{\sqrt{2}}$ with average photon number $|\alpha|^2 = \bar{n}=2$ (Fig.~3e). Because in heterodyne detection only the Q-function is directly accessible, the characteristic coherence fringes are strongly suppressed in the traveling field data; as a result, the distributions appear fairly similar. However, subtracting the marginals clearly reveals a difference, with a magnitude that is consistent with our detection efficiency and a high degree of state preservation.

The characterization of traveling fields is limited by the visibility attainable in our heterodyne detection. In particular, the averaging required for states with larger photon numbers is prohibitive. The data shown in Figs.~3c(d) have been taken with $10^7$ ($10^6$) samples per state over the course of about 12 hours. This stability of the experimental setup and the achievable detection efficiency allow us to resolve state-essential features in phase space for states containing up to $\sim 5$ photons. This lets us conclude that the release process faithfully and state-independently converts quantum states containing up to at least that size. Any imperfection in the traveling state, including state preparation or decoherence in the conversion process, can only reduce the contrast of the signatures shown. From this we estimate that the fidelities with the ideal states exceed 90\% (Supplementary Material).

The release of the cavity states shown can enable error-correctable transmission of quantum information. Because we have temporal control over the pump tones, we can shape the wave packet, which enables capture of emitted fields by a receiving module \cite{Cirac1997}\supplement. An inherent challenge for this direct quantum state transmission is that inevitable photon loss in the transmission channel will corrupt the received state. However, by choosing an appropriate encoding, the receiver will be able to detect and correct this error. For example, the states $\ket{2}$ and $(\ket{0}+\ket{4})/\sqrt{2}$ are codewords of a binomial code \cite{Michael2016} that can readily be sent by our system. Single photon loss in the transmission channel will result in a change of parity, which can be detected and correctd by a receiver.

\section*{Entanglement between stationary and traveling fields}

We next show that temporal control over the pumps allows us to generate entanglement between cavity and traveling modes by partial conversion. We use the large on/off ratio over the release process to convert only a part of the energy stored in the cavity; this is the analogue of a partially reflective beam splitter, and can thus generate entanglement between the reflected (remaining in the cavity) and transmitted field (in the transmission line). We prepare an input state in the cavity and then release half of its energy while recording the output field. This `half-release' corresponds to a 50:50 beam splitter, for which we expect maximally entangled states. After switching off the conversion process, we immediately perform a single-shot, high-fidelity ($\gtrsim 0.95$) measurement on the cavity. The non-classical correlations between recorded field and cavity outcomes measured in different bases are indicative of the generation of entanglement (Fig.~4a). We demonstrate this using two different state encodings, single photons and cat-states.

\begin{figure*}[tp]
    \center
    \includegraphics{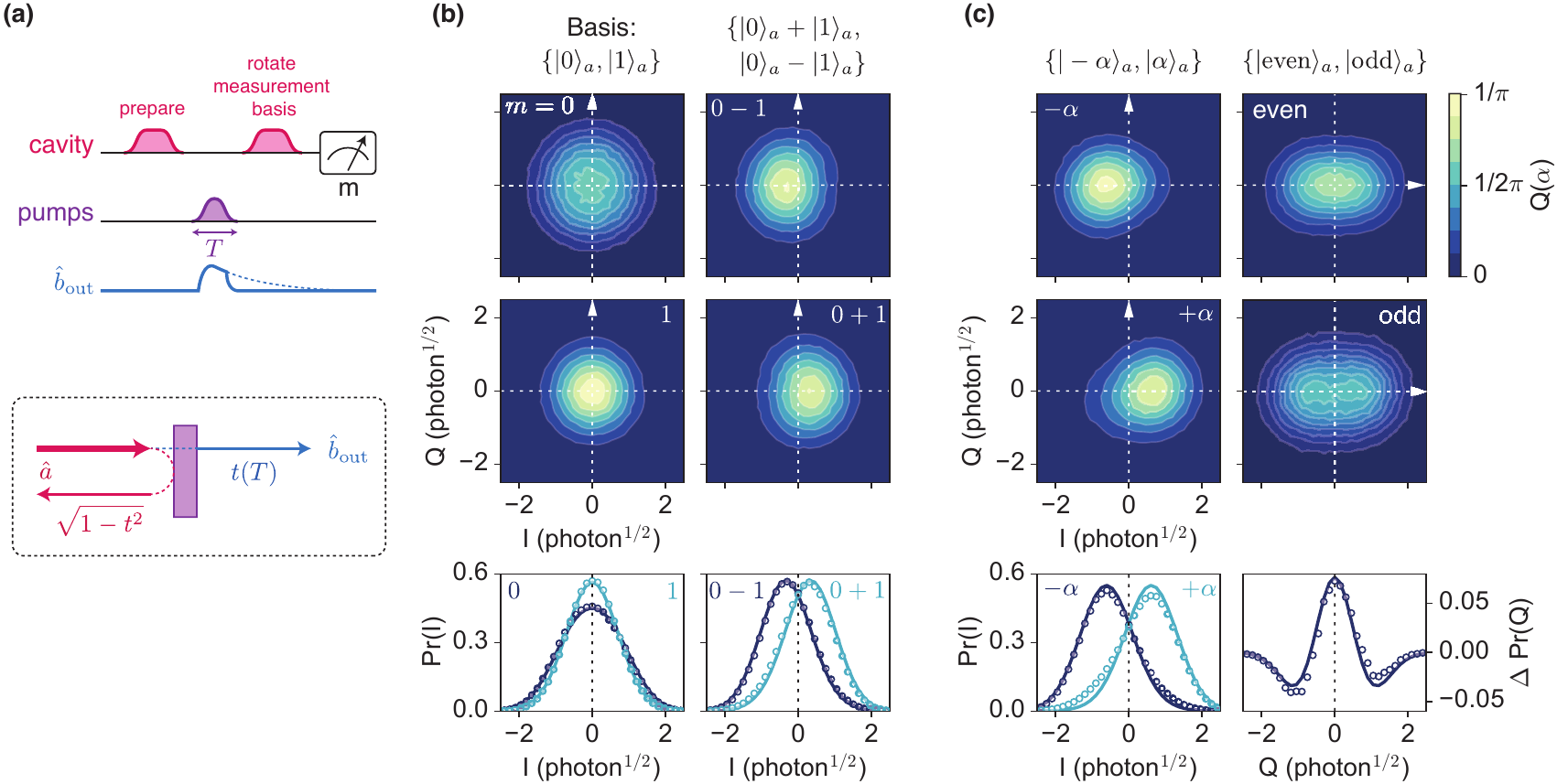}
    \caption{\label{fig:entanglement}
    Generating entanglement between stationary and traveling fields.
    {\bf a,} We partially release the cavity field by applying the pumps for a reduced amount of time. We condition the recorded field on the outcome of a subsequent measurement of the cavity to obtain correlations. Partial release is analogous to a field impinging on a partially transmitting mirror, where the transmittance $t$ is set by the pump time $T$.
    {\bf b,} Half-release of $\ket{1}$. Left column: Q-functions conditioned on finding the cavity in either $\ket{0}$ or $\ket{1}$. Right column: finding the cavity in either $\ket{0}+\ket{1}$ or $\ket{0}-\ket{1}$.
    {\bf c,} Half-release of $\ket{\mathcal{C}^{+}_{\sqrt{2}}}$. Left: conditioned on finding the cavity in either $\ket{-\alpha}$ or $\ket{+\alpha}$. Right: conditioned on finding either an even or odd number of photons in the cavity. Lower right: for the parity-conditioned data we show the difference in the $Q$-marginals. Solid lines: ideal case (perfect entangled state and perfect cavity measurement), taking into account only the detection efficiency in the Q-functions.
    All data have been taken with $g/2\pi$ = 164\,kHz and $10^6$ samples per state and basis.
    }
\end{figure*}

Half-releasing the Fock state $\ket{1}$ results in the Bell-state $(\ket{1}\ket{0} + \ket{0}\ket{1})/\sqrt{2}$, where the first ket is the state inside the cavity, and the second the traveling state. When we measure the cavity in the number basis and find it to be in the state $\ket{0}$ ($\ket{1}$), we expect to find the traveling state in $\ket{1}$ ($\ket{0}$). This is revealed with near-ideal contrast in the Q-functions from the traveling field, conditioned on cavity outcomes (Fig.~4b, left column). To show non-classicality in the correlations, we measure the cavity also in a rotated basis to probe the states $(\ket{0}\pm\ket{1})/\sqrt{2}$. Conditioning on the these outcomes, we find that the Q-functions closely resemble those of $(\ket{0}\mp\ket{1})/\sqrt{2}$, consistent with a high-fidelity entangled state. This data allows us to further confirm the presence of entanglement through a simple witness. We estimate a lower bound on fidelity compared with the ideal Bell-state of $0.91 \pm 0.02$; this clearly exceeds the classical bound of 0.5 and confirms that the half-release generates entanglement\supplement.

While entanglement with single traveling photons can also be observed with two-level systems \cite{Eichler2012}, the conversion method presented here can generate entanglement between non-classical multiphoton states. In particular, half-release of a cat state $\catket{+}{\sqrt{2}\alpha}$ generates the entangled state $\ket{\alpha}\ket{\alpha} + \ket{-\alpha}\ket{-\alpha} = \catket{+}{\alpha}\catket{+}{\alpha} + \catket{-}{\alpha}\catket{-}{\alpha}$, where we have omitted normalization factors. While such two-mode entangled cat states have been created previously in locally coupled oscillators \cite{Wang2016} and with itinerant optical photons \cite{Ourjoumtsev2009}, our scheme realizes an interface between stationary and flying cats. We demonstrate this ``Schr\"odinger catapult'' by half-releasing the cat state $\catket{+}{\sqrt{2}}$. To show non-classical correlations we measure the cavity in the coherent state basis, finding it in either $\ket{\pm\alpha}$, or in the parity basis, thus finding it in either the even or odd cat state $\catket{\pm}{1}$. The conditioned Q-functions of the flying field are shown in Fig.~4c. Again, the correlations are consistent with a high-fidelity entangled state. A slight reduction of contrast in the coherent state basis results from state evolution due to the Kerr effect, which reduces the fidelity of the cavity measurement\supplement.

This entanglement between stationary and traveling cats will enable error-correctable distribution of entanglement. Capture of the wave packet emitted by half-release enables the creation of remote entanglement between stationary parties. For the cat states used in this work, any photon loss in the transmission channel will corrupt the state because it results in change of parity. However, photon loss becomes detectable when we half-release a cat state of the form $\ket{\alpha} + \ket{\ii\alpha} + \ket{-\alpha} + \ket{-\ii\alpha}$ \cite{Roy2016}. Such states are eigenstates of `superparity' with modulo\,4 photons, and even/odd (modulo\,2) parity measurements can be used to detect and correct single-photon loss \cite{Ofek2016}. Thus, measuring and comparing the parity between the remote parties will allow for detection and correction of single-photon loss in the transmission line during remote entanglement generation.

\section*{Summary and Outlook}

We have shown the coherent release of quantum states from a microwave cavity memory. This release is enabled by parametric upconversion utilizing the non-linearity of a single Josephson junction. This conversion scheme fulfills our requirements for an interface between stationary and traveling oscillator states in a microwave quantum network: We can dynamically control the conversion rate, releasing cavity states almost 1000 times faster than the intrinsic life time. This conversion rate is state-independent for states containing up to a few photons, extendable by simple hardware adjustments. The release process is equivalent to a beam splitter interaction, and cavity states are mapped faithfully onto traveling states. This interaction can be controlled precisely and rapidly, enabling the generation of entanglement between cavity and traveling modes.

Our interface can serve as the backbone in a microwave quantum network in which quantum information is stored in cavities. Since the conversion process is controllable in amplitude and phase by the pump tones, it can be used also for capturing traveling states\supplement. Combining multiple copies of our system will thus allow quantum state transfer and entanglement between remote cavities \cite{Cirac1997}. The scheme presented supports multidimensional Hilbert spaces, thus providing a route towards error-correctable distribution of quantum information and entanglement. Loss-resilient quantum communication between isolated modules will be crucial for scaling up future quantum information processing devices.

The on-demand generation of arbitrary, traveling multiphoton quantum states shown will enable exciting new opportunities for hybrid quantum systems. For instance, while mechanical oscillators can be highly coherent, it is difficult to create non-classical states by coupling them directly to nonlinear systems. However, the efficient capture of traveling microwave fields has been demonstrated experimentally \cite{Andrews2015}. Integration with our `catapult' will thus enable the creation of highly non-classical mechanical states. Mechanical systems can act as transducers with radically different degrees of freedom, such as light in the optical domain \cite{Bochmann2013,Andrews2014}. The integration of our system with such a transducer will thus enable the distribution of exotic continuous variable quantum states in heterogeneous networks.

\section*{Acknowledgements}
We thank J.~Blumoff, K.~Chou, M.~Constantin, and M.~Reagor for experimental assistance, and K.~Sliwa and A.~Narla for help setting up the parametric amplifier. We gratefully acknowledge valuable discussions with S.M.~Girvin, R.~Hanson, Z.~Leghtas, K.W.~Lehnert, A.~Narla, A.~Reed, S.~Shankar, and S.~Touzard.
This research was supported by the U.S. Army Research Office (W911NF-14-1-0011). W.P.\ was supported by NSF grant PHY1309996 and by a fellowship instituted with a Max Planck Research Award from the Alexander von Humboldt Foundation; W.P.\ and P.R.\ by the U.S.\ Air Force Office of Scientific Research (FA9550-15-1-0015); C.J.A.\ by an NSF Graduate Research Fellowship (DGE-1122492); L.J.\ by the Alfred P.\ Sloan Foundation and the Packard Foundation.
Facilities use was supported by the Yale Institute for Nanoscience and Quantum Engineering (YINQE), the Yale SEAS cleanroom, and the National Science Foundation (MRSECDMR-1119826).

\appendix

\renewcommand{\theequation}{S\arabic{equation}}
\renewcommand{\thefigure}{S\arabic{figure}}
\renewcommand{\thetable}{S\arabic{table}}
\setcounter{equation}{0}
\setcounter{figure}{0}
\makeatletter


\section{Experimental system}

\subsection{Experiment set-up}

\begin{figure*}[htp]
    \center
    \includegraphics{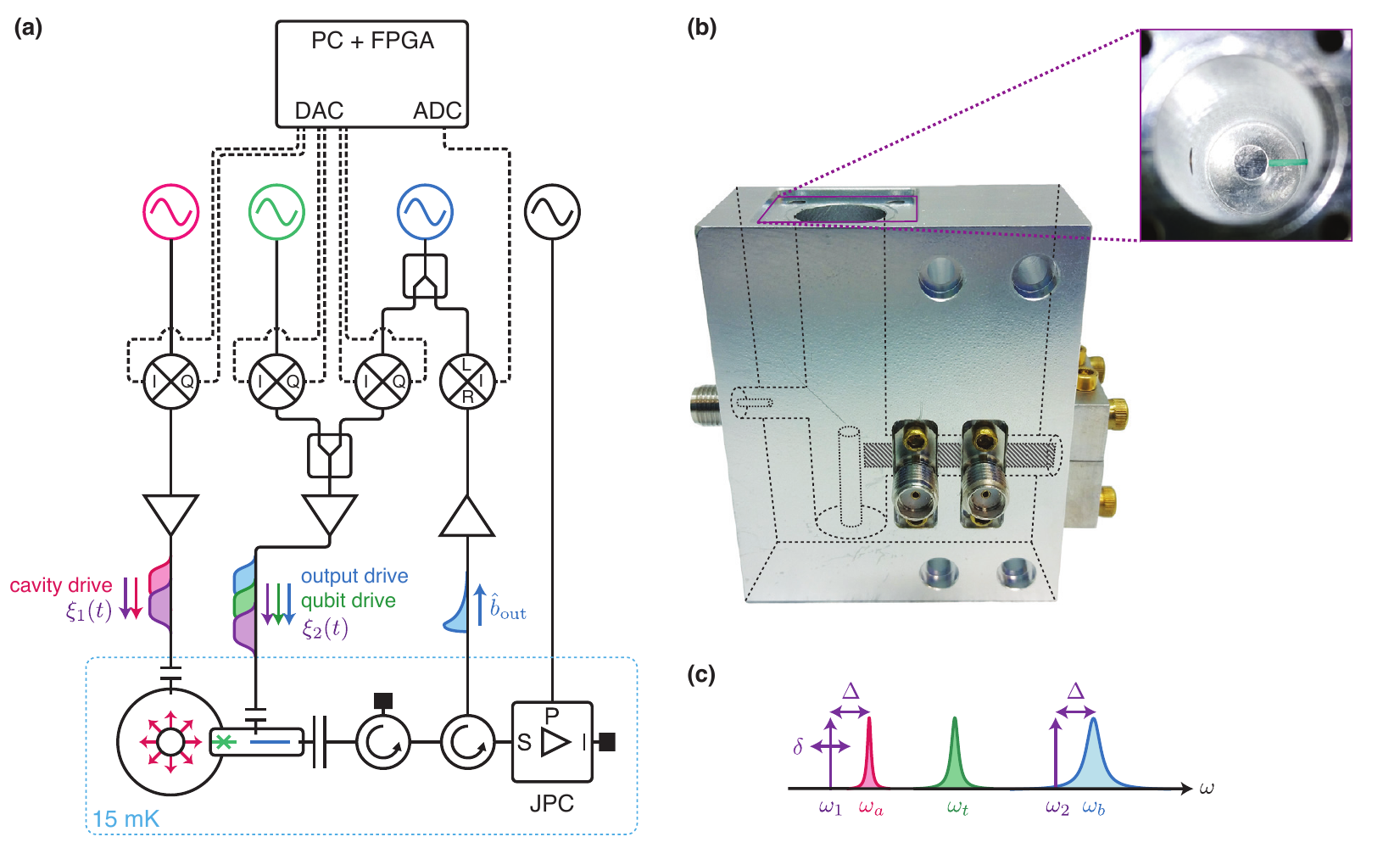}
    \caption{\label{fig:expt_scheme}
    Experimental setup.
    {\bf (a)} Wiring diagram (schematic). Each mode has a microwave generator as local oscillator (LO) (cavity and output: {\em Agilent E8275D}; transmon: {\em Vaunix LabBrick LMS-103-13}). IQ modulation tones are synthesized by an integrated FPGA system with digital-to-analog converter (DAC) outputs ({\em Innovative Integration VPXI-ePC}), and mixed with the LO ({\em Marki Microwave IQ0618LXP} IQ-mixers). Signals are amplified at room temperature ({\em MiniCircuits ZVA-183-S+}) and sent into the refrigerator ({\em Oxford Triton 200}) where the sample is cooled to $T \approx 15\,\text{mK}$. Input signals are applied through weakly coupled pins (depicted as small capacitor). Signals leave the output resonator through a strongly coupled pin (depicted as large capacitor) and is amplified by a JPC that is pumped continuously by a microwave generator ({\em Agilent N5183A}). The signal is further amplified at 4\,K ({\em Caltech CIT1-4254-065}) and at room temperature ({\em Miteq AFS3-00101200-35-ULN}), mixed down with the output LO ({\em Marki Microwave IQ0618LXP}), and recorded and demodulated by the FPGA system via analog-to-digital converters (ADC).
    {\bf (b)} Sample. The location of the stub cavity as well as the coupler and chip tunnels are indicated by dashed lines; the position of the sapphire chip is indicated by the hatched rectangle. Cavity signals are applied through the SMA connector visible on the left, output mode signals enter and leave through the connectors at the front. The chip is held in place by a clamp on the right side. The inset shows a top-down view into the cavity, with the inserted chip false-colored in green.
    {\bf (c)} Frequency ordering of the modes. The pumps are applied red-detuned from the output and storage modes, with common detuning $\Delta = 2\pi \times (30-50)\,\text{MHz}$. An additional `relative detuning', $\delta$, between the pumps allows for compensation of Stark shifts and to study the behavior if the conversion Hamiltonian is off-resonant.
    }
\end{figure*}

\paragraph*{Setup and signals}
The sample is cooled to $T \approx 15\,\text{mK}$ in a dilution refrigerator. A wiring diagram showing how signals are introduced to the device is shown in Figure \ref{fig:expt_scheme}a. Each mode of the system is addressed by a separate microwave generator acting as a local oscillator (LO); pulses are generated by IQ modulation. Importantly, we generate the pump tones using the same generators used for the (near-)resonant control pulses of the cavity and output mode. Given the conversion Hamiltonian, this guarantees that the signal emitted from the output mode is phase-locked to the output mode LO, which is used for mixing the signal down to low frequencies before digitizing.

\paragraph*{Quantum-limited amplification}
The output signal is processed by a Josephson parametric converter (JPC) operated in amplification mode \cite{Bergeal2010}. The amplifier is configured to provide approximately 25\,dB of gain with a bandwidth of approximately 15\,MHz and a noise visibility ratio around 6\,dB. This allows us to detect signals emitted at the frequency of the output mode with a detection efficiency of $\sim 45\,\%$ (details follow).


\subsection{Sample}

\paragraph*{Device description}
The device used to perform this experiment includes a high-Q 3D coaxial stub cavity \cite{Reagor2016}, transmon qubit, and output resonator within a single package \cite{Axline2016}. The package is machined from high-purity (4N) aluminum to allow for long single-photon lifetimes in the cavity. The transmon qubit and output resonator are fabricated on a single sapphire chip that is inserted into a tunnel and positioned in close enough proximity to interact strongly with the 3D cavity (Figure \ref{fig:expt_scheme}b). Parameters of each mode, including frequencies and coherence times, are included in Table \ref{tab:devparams}.

\paragraph*{System Hamiltonian}
The Hamiltonian of the system consisting of the cavity ($a$), output resonator ($b$), and transmon ($c$) is, in the absence of drives, given by \cite{Nigg2012}
\begin{equation}
    \label{eq:undriven_H}
    H/\hbar = \omega_a \modea^\dagger\modea + \omega_b \modeb^\dagger\modeb
        + \omega_c \modet^\dagger\modet - E_\text{J}\left(\cos (\hat{\varphi})
        - \frac{\hat{\varphi}^2}{2} \right).
\end{equation}
The flux across the junction is given by
\begin{equation}
    \hat{\varphi} = \sum_{k=a,b,c} \phi_k (\mode{k} + \mode{k}^\dagger),
\end{equation}
where $\phi_k$ are magnitudes of the zero-point fluctuations of flux across the junction. Transforming into the rotating frame of all modes and expanding the cosine to $\hat{\varphi}^4$ gives
\begin{equation}
    H/\hbar = \sum_{k\neq l} \chi_{kl} \mode{k}^\dagger \mode{k} \mode{l}^\dagger \mode{l}
        + \sum_{k} \frac{\chi_{kk}}{2} \mode{k}^{\dagger 2} \mode{k}^2,
\end{equation}
where we have introduced the cross-Kerr coefficient (or dispersive shift) $\chi_{kl} = -E_\text{J}\phi_k^2\phi_l^2$ and the self-Kerr coefficient (or anharmonicity) $\chi_{kk} = -E_\text{J}\phi_k^4/2$. Measured mode frequencies and nonlinearities up to this order are listed in Table~\ref{tab:devparams}. Details on how these parameters can be measured can be found, for instance, in refs.~\cite{Kirchmair2013,Vlastakis2013}.

\begin{table}[tb]
    \centering
    \begin{tabular}{l c c} 
    \hline\hline
    Hamiltonian term &  & Value (MHz) \\
    \hline
    Mode frequency    & $\omega_{b}/2\pi$     & $9.999 \times 10^3$  \\
                      & $\omega_{a}/2\pi$	  & $4.073 \times 10^3$ \\
                      & $\omega_{c}/2\pi$	  & $6.674 \times 10^3$ \\
    Cross-Kerr        & $\chi_{\text{ab}}/2\pi$     &  $-0.013 \pm 0.001$ \\
                      & $\chi_{\text{ac}}/2\pi$     &  $-3.825 \pm 0.001$ \\
                      & $\chi_{\text{bc}}/2\pi$     &  $-1.3 \pm 0.1$ \\
    Self-Kerr         & $\chi_{\text{aa}}/2\pi$     &  $-0.022 \pm 0.002$ \\
                      & $\chi_{\text{bb}}/2\pi$     &  $-0.001 \pm 0.001$ \\
                      & $\chi_{\text{cc}}/2\pi$     &  $-144.0 \pm 0.5$ \\
    \hline
    Damping term & & Value ($\mu$s) \\
    \hline
    Cavity energy decay       & $1/\kappa_{0}$ & $450 \pm 50$  \\
    Output energy decay       & $1/\kappa_{\text{out}}$ & $0.24 \pm 0.02$\\
    Transmon qubit relaxation & $T_{1}$  			  & $50 \pm 10$ \\
    Transmon Ramsey decay     & $T_{2R}$	     	  & $25 \pm 5$ \\
    Transmon Hahn echo decay  & $T_{2E}$			  & $35 \pm 5$ \\
    \hline\hline
    \end{tabular}

    \caption{
    \label{tab:devparams}
    Measured system parameters. See text for explanation. For the cavity and transmon qubit decay times, the uncertainties given are the typical fluctuations observed in the course of a day.}
\end{table}

\subsection{System preparation and readout}

\paragraph*{Measurement of the transmon}
The dispersive coupling between the transmon and output resonator, $\chi_{bc}\modeb^\dagger\modeb \modet^\dagger \modet$, allows high-fidelity measurement of the transmon qubit state. Because the frequency of the output depends on the transmon state, the phase of a transmitted signal differs when the transmon is in the ground ($\ket{g}$) or excited state ($\ket{e}$). Measuring this phase thus corresponds to a measurement of the transmon state. A $2.5\,\mu\text{s}$ readout pulse with an average of $\bar{n} \approx 5$ photons allows us to discriminate between the two states with a fidelity $> 0.99$. In the absence of decay-events of the transmon, this contrast represents the measurement fidelity. However, the effect of transmon $T_1$ causes measurement outcomes to be incorrectly assigned to the $\ket{g}$ state; this reduces the fidelity with which we can assign $\ket{g}$ to 0.96.

\paragraph*{Measurement of the storage cavity}
Dispersive coupling between transmon and storage cavity allows readout of the cavity state. We can probe whether the cavity contains $n$ photons by first applying a $\pi$-pulse, selective of the transition for the cavity containing $n$ photons, followed by a measurement of the transmon state. The fidelity of the positive readout result (cavity containing $n$ photons) is simply the transmon readout fidelity for $\ket{e}$, $\sim 0.99$. The negative result (cavity not containing $n$ photons) is determined by the fidelity for $\ket{g}$ and the probability with which the $n$-photon state has not been mapped sucessfully onto $\ket{e}$; the combined fidelity for the negative results is about $0.9$ and is limited by qubit decoherence.

\paragraph*{Feedback cooling}
In equilibrium we find a finite thermal population of the qubit and cavity; the qubit population in $\ket{e}$ is $\sim 0.08$ and the cavity has an average number of thermal photons of $\bar{n} \approx 0.1$. We cool the system through a sequence of QND measurements and selective $\pi$-pulses on the transmon \cite{Heeres2016}. We achieve a qubit population in $\ket{e}$ of $\sim 0.02$, and a cavity population of $< 0.01$ before the start of each measurement sequence.
\paragraph*{State creation}
To prepare complex cavity states, we apply pulses at qubit and cavity frequencies that are designed with a numerically optimized control technique \cite{Heeres2016}. This state preparation technique produces the cavity states used in this experiment within 1 $\mu$s and with typical fidelities above 95\,\%.

\section{Model of the conversion from cavity to traveling mode}
\label{sec:mode-conversion}

\paragraph*{Full driven Hamiltonian}
Taking into account the pump drives applied to the output and cavity, the full Hamiltonian is
\begin{equation}
\begin{aligned}
    \label{eq:driven_H}
    H/\hbar =\, & \omega_a \modea^\dagger\modea + \omega_b \modeb^\dagger\modeb
        + \omega_c \modet^\dagger\modet - E_\text{J} \left( \cos (\hat{\varphi})
        - \frac{\hat{\varphi}^2}{2} \right)\\
        & + \epsilon_1(t)\expe^{-\ii \omega_1 t} (\modea^\dagger + \modea)\\
        & + \epsilon_2(t)\expe^{-\ii \omega_2 t} (\modeb^\dagger + \modeb),
\end{aligned}
\end{equation}
where $\epsilon_{1,2}$ are the drive amplitudes. Moving into a displaced, rotating frame we can rewrite this as \cite{Leghtas2015}
\begin{equation}
\begin{aligned}
    H/\hbar = - E_\text{J} \cos\Big(&
        \phi_a(\rotmode{a}^\dagger + \rotmode{a} + \tilde\xi_1^*(t) + \tilde\xi_1(t)) \\
        &+ \phi_b(\rotmode{b}^\dagger + \rotmode{b} + \tilde\xi_2^*(t) + \tilde\xi_2(t)) \\
        & + \phi_c(\rotmode{c}^\dagger + \rotmode{c})
    \Big) - E_\text{J} \frac{\hat{\varphi}^2}{2}.
\end{aligned}
\end{equation}
We have now introduced the rotating field operators $\rotmode{k} = \mode{k}\exp(-\ii \omega_k t)$ and the classical displacement amplitudes
\begin{equation}
    \tilde\xi_i (t) =  \xi_i(t)\expe^{-\ii \omega_i t} = \frac{\epsilon_i(t) \expe^{-\ii \omega_i t}}{\ii \kappa_i/4 + \Delta_i}
        \approx \frac{\epsilon_i(t) \expe^{-\ii \omega_i t}}{\Delta_i}.
\end{equation}
Here, we have separated the slowly varying envelope of the displacement, $\xi_i(t)$, from its fast oscillating component. The detuning $\Delta_i$ is given by the detuning of the drive frequency with respect to the mode that the drive is applied to, i.e., $\Delta_1 = \omega_1 - \omega_a$ and $\Delta_2 = \omega_2 - \omega_b$. $\kappa_i$ is the damping rate of the mode to which the pump is applied.

\paragraph*{Conversion Hamiltonian}
We make the rotating wave approximation (RWA) by expanding the cosine up to orders of $\phi_k^4$ and keeping all terms that conserve energy. Besides the terms that are already present in the undriven case, we now find the additional terms
\begin{equation}
\begin{aligned}
    H_1/\hbar = & |\xi_1|^2 \left(
            2\chi_{aa} \modea^\dagger \modea +
                \chi_{ab} \modeb^\dagger \modeb + \chi_{ac} \modet^\dagger \modet
        \right) \\
        & + |\xi_2|^2 \left(
            2\chi_{bb} \modeb^\dagger \modeb +
                \chi_{ab} \modea^\dagger \modea + \chi_{bc} \modet^\dagger \modet
        \right) \\
        & + g \modea \modeb^\dagger \expe^{-\ii \delta t}
            + g^* \modea^\dagger \modeb \expe^{\ii \delta t}.
\end{aligned}\label{eq:stark_hamiltonian}
\end{equation}
The first two lines in this expression correspond to Stark shifts, for which we can compensate by detuning the drives. The last line is the desired conversion Hamiltonian, where $g = \chi_{ab}\xi_1^* \xi_2$. $\delta = \Delta_1 - \Delta_2$ denotes a (small) detuning from the conversion process. We have assumed that the pumps are tuned such that the conversion process is close to resonant, and that all other terms are fast-oscillating and vanish within the RWA.

\paragraph*{Emitted field}
Using input-output theory \cite{Steck2015}, we can write the coupled equations of motion for the system,
\begin{subequations}
\label{eq:equationsofmotion}
\begin{align}
    \deriv{t} \mode{a} &= -\ii \commute{\mode{a}}{H} = -\ii g \mode{b}, \\
    \deriv{t} \mode{b} &= -\ii \commute{\mode{b}}{H} - \frac{\kappaout}{2} \mode{b}
        = -\ii g \mode{a} - \frac{\kappaout}{2} \mode{b}, \\
    \modebout &= \sqrt{\kappaout} \modeb.
\end{align}
\end{subequations}
For the Hamiltonian we only consider the conversion term. We have ignored any loss aside from coupling to the transmission line, and we ignore input modes $\modea_\text{in}, \modeb_{in}$ because they are in the vacuum state during the release. For simplicity, we take $g$ to be real, but it can be shown that the phase of this coupling term is simply mapped onto the output field. These equations can be solved analytically, assuming some initial condition $\mode{a}(0)$:
\begin{subequations}
\begin{align}
	\mode{a}(t) &= \frac{\mode{a}(0)}{\beta} \expe^{-\frac{\gamma t}{4}}
    	\left(
        	\beta\cosh\left(\frac{t\beta}{4}\right) +
            \gamma\sinh\left(\frac{t\beta}{4}\right)
        \right),\\
    \mode{b}(t) &= -\ii \frac{4g\mode{a}(0)}{\beta}
    	\expe^{-\frac{t}{4}(\kappaout - 2\ii\delta)} \sinh\left(\frac{t\beta}{4}\right),
\end{align}
\end{subequations}
where we have introduced
\begin{subequations}
\begin{align}
	\gamma &= \kappaout + 2\ii\delta,\\
    \beta &= \sqrt{\gamma^2 - (4g)^2}.
\end{align}
\end{subequations}
We first consider the case $\delta=0$, i.e., the conversion Hamiltonian is resonant. We are mostly concerned with the case $g \ll \kappaout$; in this regime the coupling results in exponential damping of the cavity, which can be seen by approximating
\begin{subequations}
\begin{align}
    \mode{a}(t) &\approx \mode{a}(0) \expe^{-2g^2t/\kappaout}, \\
    \mode{b}(t) &\approx -\ii \mode{a}(0) \frac{2g}{\kappaout}
        \left( \expe^{-\kappa t/2} - \expe^{-\kappaout t/2} \right), \\
    \mode{b}_\text{out}(t) &= \sqrt{\kappaout}\mode{b}(t)
        \approx -\ii \frac{\kappa}{2}\mode{a}(t),
\end{align}
\end{subequations}
where $\kappa = 4g^2/\kappaout$. We note that this is equivalent to the Purcell (spontaneous emission) rate for a resonant Jaynes-Cummings interaction enabled by the pumps.

\paragraph*{Effect of detuning}
For $\delta \neq 0$ we can approximate $\beta \approx \gamma - 8g^2/\gamma$. The evolution of $\mode{a}$ is then
\begin{align}
	\mode{a}(t) &\approx \mode{a}(0) \expe^{-\frac{t}{2}\frac{4g^2}{\gamma}} \equiv \mode{a}(0) \expe^{-\frac{t}{2} \gamma_\text{conv}}.
\end{align}
Since $\gamma_\text{conv}$ is complex, we must evaluate its real part to obtain the linewidth of the conversion resonance. We have
\begin{equation}
	\gamma_\text{conv} = \frac{4g^2}{\kappaout + 2\ii\delta} =
    	\frac{4g^2(\kappaout - 2\ii\delta)}{\kappaout^2 + (2\delta)^2},
\end{equation}
the real part of which is a Lorentzian with full-width half-maximum $\kappaout$ as a function of detuning. Figure \ref{fig:qswitch-details}a shows the decay of the cavity versus detuning, confirming this expectation.

\paragraph*{Beam splitter description}
The release from  the storage cavity into the transmission line mode can also be described by an effective beam splitter interaction, which does not explicitly take the output mode into account. This interaction is described by the unitary transformation
\begin{equation}
    \label{eq:bs-a-bout}
    U_\text{release} = \exp\left(\frac{\theta}{2} (\modea^\dagger \modebout - \modea \modebout^\dagger)\right)
\end{equation}
that acts on the cavity and propagating mode. For $g\ll\kappaout$, the cavity decays exponentially with a rate $\kappa = 4g^2/\kappaout$. In that case, the mixing angle $\theta$ is given by
\begin{equation}
    \label{eq:bs-a-bout-theta}
    \theta = 2 \arccos\left( \exp(-\kappa t/2)\right).
\end{equation}
For example, a 50:50 beam splitter interaction can then be achieved by a pump duration $t = \ln2/\kappa$, corresponding with $\theta = \pi/2$.

\section{Supplementary methods and analysis}

\subsection{Coupling strength calibration}

\begin{figure}[tp]
    \center
    \includegraphics{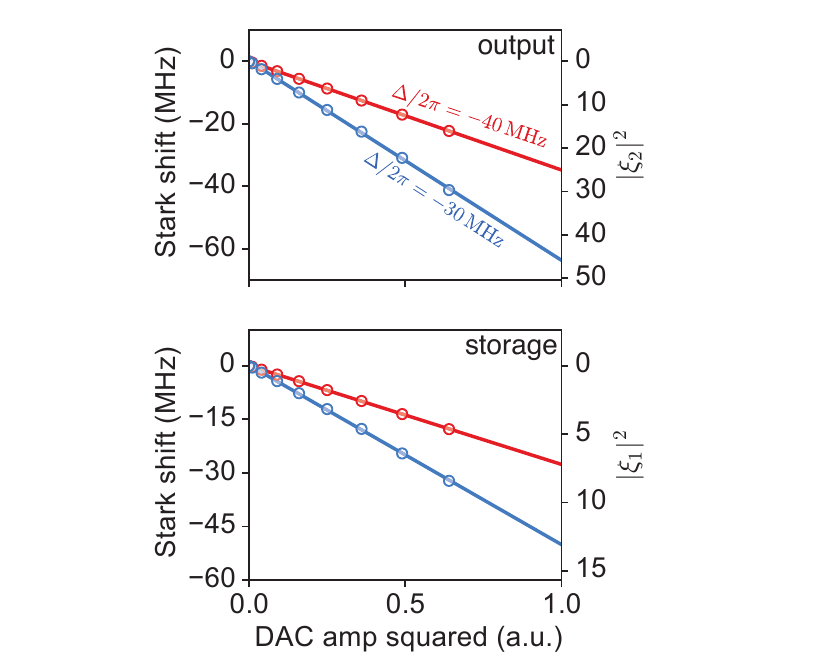}
    \caption{\label{fig:stark_shift}
    Pump strength calibration using the transmon's Stark shift. Applying pump tones, controlled by a DAC voltage, on output (upper panel) and storage mode (lower panel) results in a Stark shift of the transmon mode, measured by spectroscopy while applying the pump. Since the Stark shift is proportional to the photon number in the pump (which is proportional to the drive amplitude squared), we obtain a calibration for $\xi_{1,2}$. Solid lines are linear fits.
    }
\end{figure}

The conversion rate is given by the product of the dispersive shift between storage and output modes and the pump strength, $g = \chi_{ab} \xi^*_1 \xi_2$. To calibrate this rate, we need to determine the number of pump photons $|\xi_{1(2)}|^2$ in the storage (output) mode when applying the drives. This can be done by measuring the Stark shift of the transmon mode: when applying the pump tone on the storage (output) resonator, the Stark shift is given by $\delta\omega = \chi_{ac(bc)}|\xi_{1(2)}|^2$. Since $\chi_{ac}$ and $\chi_{bc}$ can be determined independently, measuring the Stark shift is a calibration for $\xi_{1},\xi_{2}$. For $\Delta/2\pi = -30$\,MHz and $-40$\,MHz we show this calibration in Fig.~\ref{fig:stark_shift}. We then use this pump calibration to obtain $\chi_{ab}$. Applying a pump tone on the output mode $\modeb$ results in a Stark shift of mode $\modea$ with magnitude $\chi_{ab} |\xi_2|^2$. Since $\xi_2$ is known, measuring the Stark shift yields $\chi_{ab}$, and the value of $g$ can then be calculated.

\subsection{Cavity Q-switch}

\begin{figure}[tp]
    \center
    \includegraphics{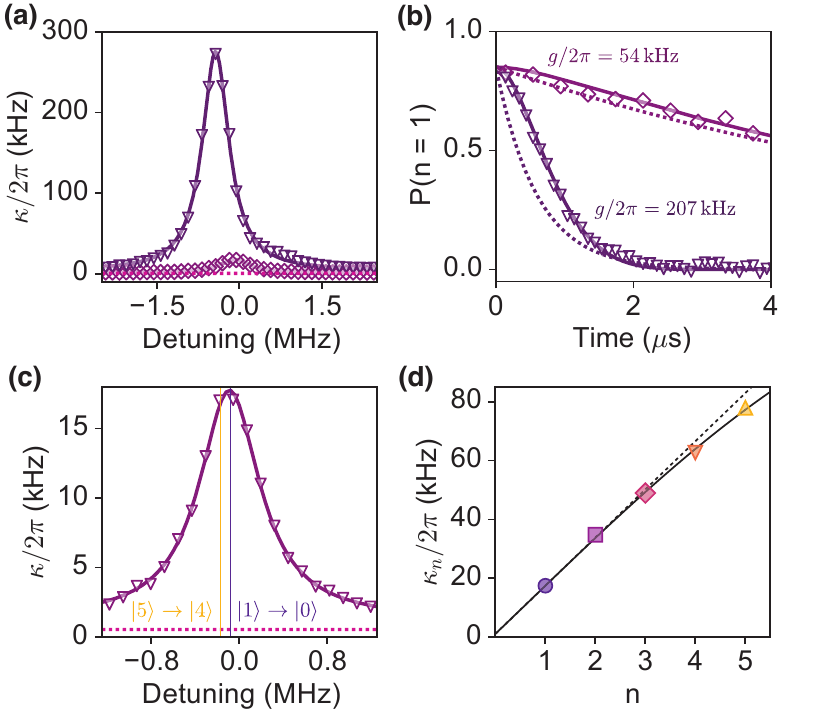}
    \caption{\label{fig:qswitch-details}
    Cavity damping analysis.
    {\bf (a)} Cavity damping as function of relative pump detuning $\delta$. The two data sets represent $g/2\pi =$ 54 and 207\,kHz. Each point is obtained by fitting individual decay curves with a single exponential, $\propto \exp(-\kappa t)$. Solid lines are Lorentzian fits. The resonance condition for the conversion process is offset from 0 because the pumps induce a Stark shift on the resonators. Note that for the larger $g$ the decay rate $\kappa$ is not a very good approximation because the damping is not purely exponential in time.
    {\bf (b)} Non-exponential decay for large $g$. Data is the same as in Fig.~2a of the main text, with pump frequencies on resonance with the conversion process. Solid lines: complete model, based on independent calibration of $g$. Dashed line: $4g^2/\kappaout$ approximation.
    {\bf (c)} Reduction of $\kappa_n$ due to the Kerr effect. Data and fit are for $g/2\pi =$ 54\,kHz, the setting with which the data shown in Fig.~2c of the main text have been obtained. Orange and blue lines mark the resonance condition for one and five photon Fock states, respectively.
    {\bf (d)} Decay of the Fock states $\ket{n}$. Dashed line: extrapolation from $\kappa_1$. Solid line: Correction that takes Kerr and the Lorentzian profile of the resonance into account. The deviation between the measured $\kappa_5$ and $5 \times \kappa_1$ is 6\%.
    }
\end{figure}

\paragraph*{Detuning}
When the four-wave mixing process is detuned (as described above), the conversion Hamiltonian becomes
\begin{equation}
    H_\text{conv} = g \modea \modeb^\dagger \expe^{-\ii \delta t} + \text{h.c.},
\end{equation}
where $\delta$ is the detuning from the process. Ignoring Stark shifts, this situation arises if one pump (for example, $\xi_1$) is detuned $\Delta$ from its respective mode resonance (e.g. to $\omega_a+\Delta$), and the other by $\Delta + \delta$ (in our example, $\xi_2$ is detuned to $\omega_b+\Delta+\delta$). For $g \ll \kappaout$ the decay constant of the storage mode is then given by a Lorentzian of width $\kappaout$,
\begin{equation}
    \kappa(\delta) = \frac{4g^2 \kappaout}{\kappaout^2 + 4\delta^2}.
\end{equation}
We measure $\kappa(\delta)$ by fixing $\Delta$ for one of the pumps and sweeping the detuning $\delta$, thus affecting only one pump frequency. At each point we measure the decay of the storage mode and extract $\kappa$ via a single exponential fit. The data for two different values of $g$ are shown in Fig.~\ref{fig:qswitch-details}a.

\paragraph*{Expected conversion efficiency}
The excellent agreement of the data with a Lorentzian line shape indicates a resonant process. If this process is indeed the predicted conversion, we expect that its damping rate is given by the difference between the maximum of the Lorentzian fit and its offset, $\kappa - \kappa_\text{ofs}$. $\kappa_\text{ofs} = \kappa_\text{loss}$ is the damping due to any other losses, intrinsic or pump-induced. Hence, we expect that the inefficiency of the conversion is given by $\sim \kappa_\text{loss}/\kappa$. For $g/2\pi \agt 100\,\text{kHz}$ we find $\kappa_\text{loss}/\kappa$ between 0.01 and 0.015.

\paragraph*{Output mode bandwidth}
If the condition $g \ll \kappaout$ is not fulfilled, the decay of the storage mode is no longer exponential and can no longer be described accurately by a single decay rate $\kappa$. This can be seen in Fig.~\ref{fig:qswitch-details}b, where we show the damping of the storage mode for $g/2\pi =$ 54\,kHz and 207\,kHz. At 207\,kHz there is a clear deviation between the exact model and the approximation.

\paragraph*{State dependence}
We expect that for larger Fock states $\ket{n}$ the damping rate becomes slower: due to the Kerr effect in the storage mode (here, $\chi_{aa}/2\pi =$ 22\,kHz) the transition $\ket{n+1}\rightarrow\ket{n}$ is shifted by $n \chi_{aa}$. The finite bandwidth of the conversion process, $\kappaout$, means that for large $n$ the conversion becomes off-resonant, leading to a reduced $\kappa_n$ (Fig.~\ref{fig:qswitch-details}c). For $n \leq 5$ this effect is small, with a deviation of $\kappa_n$ from $n \kappa$ of $\leq$ 6\% (Fig.~\ref{fig:qswitch-details}d). For future experiments it is desirable to improve the state-independence by reducing the magnitude of the Kerr effect in the storage mode and increasing the bandwidth by a larger $\kappaout$.

\begin{figure}[tp]
    \center
    \includegraphics{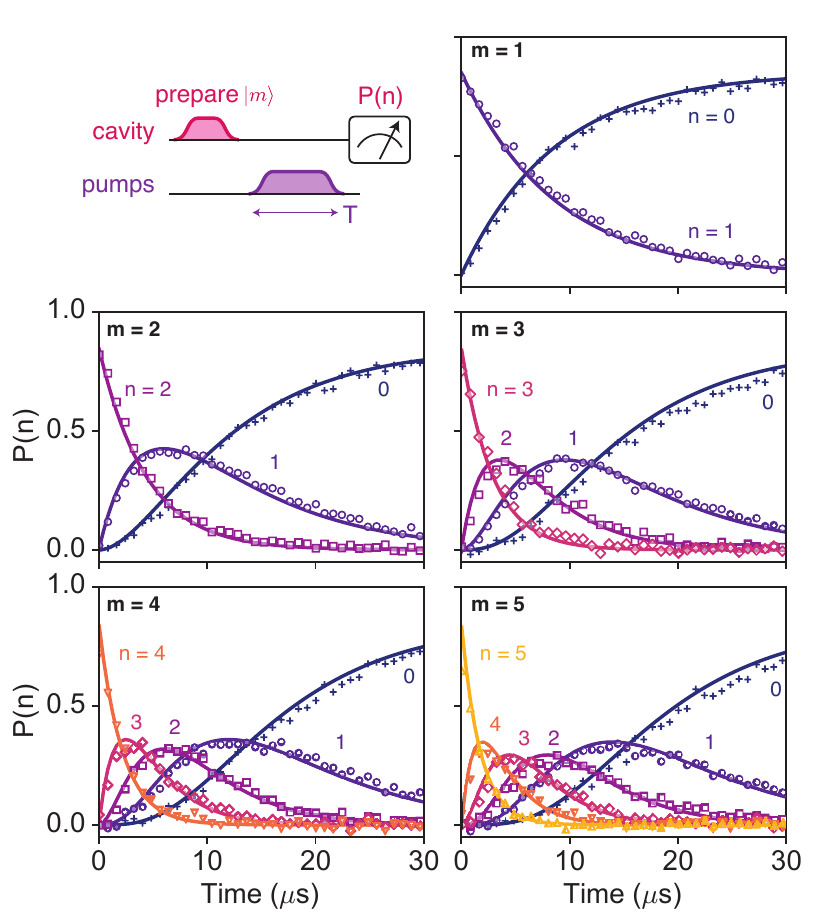}
    \caption{\label{fig:fock-evolution}
    Cavity evolution during the decay of Fock states. We measure the number state populations $P(n)$ after preparing the Fock state $\ket{m}$. Solid lines: theoretical prediction, based on independent calibration of $g$. Here, $g/2\pi =$ 54\,kHz.
    }
\end{figure}

\paragraph*{Cavity evolution during Fock state decay}
To further ensure that the release is well-described by a state-independent beam splitter, we prepare the Fock state $\ket{m}$ and measure the time-dependent cavity number state populations $P(n)(t)$, for $n\leq m$ (Fig.~\ref{fig:fock-evolution}). For purely exponential damping of the cavity (i.e., $g \ll \kappaout$ and $\kappa = 4g^2/\kappaout$), we expect
\begin{equation}
    P(n|m)(t) = \binom{m}{n} \expe^{-m\kappa t} \left( \expe^{\kappa t} - 1\right)^{m-n}.
\end{equation}
Using a conversion strength that fulfills this condition, $g/2\pi =$ 54\,kHz, we find excellent agreement between the data and the model (allowing no free parameters).

\subsection{Calibration of traveling signals}

To quantitatively analyze the propagating states, we need to determine how recorded signals relate to the original cavity state. This relation is governed by a) the efficiency $\eta_\text{conv}$ with which cavity states are converted to the output resonator, and b) the efficiency $\eta_\text{det}$ with which fields emitted from the output resonator are recorded by the detector; because the damping rate of the output mode is set by the coupling to the transmission line, we assume that all photons in the output mode are converted to traveling signals. We treat $\eta_\text{det}$ as a property of our detection circuitry, whereas $\eta_\text{conv}$ impacts the fidelity with which cavity states are launched. In the following we outline how we separate the two quantities by calibration measurements. We further lay out how we compute the probability distribution in phase space that we use to analyze traveling quantum states.

\subsubsection*{Conversion efficiency}

As noted above, the Q-switch data suggests that the conversion efficiency is high. To directly measure this efficiency, we aim to determine the number of photons contained in the wavepacket emitted while launching a known state from the cavity. To this end we first calibrate the signals recorded by the ADC in terms of photon flux emitted by the output mode, and then use this calibration to determine the number of photons contained in the launched field.

\paragraph*{ADC signal calibration}
We determine the output resonator drive amplitude that generates a steady-state coherent state with $\bar{n}_\text{ref}=1\pm0.09$ photons in the output mode via the Stark shift induced on the transmon \cite{Schuster2005}. This reference drive results in a constant emitted photon flux of $\modebout^\dagger(t)\modebout(t) = \bar{n}_\text{ref}\kappaout$, and a reference field $\langle\modebout(t)\rangle_\text{ref} = \sqrt{\kappaout\bar{n}_\text{ref}}$. Because in heterodyne detection we cannot directly measure $\langle \modebout^\dagger(t)\modebout(t) \rangle$ we instead evaluate $\langle\modebout(t)\rangle_\text{ref}^2$; for a coherent state these two expressions are identical. The number of photons is then $\int_0^{T} \langle b_\text{out}(t)\rangle_\text{ref}^2\,\dif{t} = \bar{n}_\text{ref}$, where the integration is performed up to $T = 1/\kappaout$. Measuring the emitted field while applying this steady-state drive is therefore a calibration that relates the emitted number of photons in a signal to the heterodyne voltage recorded by the ADC.

\paragraph*{Photons emitted during cavity state release}
To assess the conversion efficiency we launch the cavity state $(\ket{0}+\ket{1})/\sqrt{2}$ as the signal, and record the emitted field. For a traveling state of that form $\langle \modebout(t) \rangle^2_\text{sig} = \langle \modebout^\dagger(t)\modebout(t) \rangle /2$. In absence of conversion loss, this signal is then related to that of the reference,
\begin{equation}
    \int \langle \modebout(t) \rangle^2_\text{sig} \dif{t} = \frac{1}{4} \int_0^{T} \langle b_\text{out}(t)\rangle_\text{ref}^2\,\dif{t},
\end{equation}
where the integral over the signal is performed over the full temporal extent of the wavepacket. From this experiment we find that this signal contains $0.50 \pm 0.08$ photons for large conversion rates, $g/2\pi \agt 150\,\text{kHz}$. This value is consistent with our observation of small $\kappa_\text{loss}/\kappa$, from which we expected a conversion efficiency of $\etaconv \approx 0.98-0.99$. We note that this measurement of the conversion efficiency is independent of the detector, because both the reference and the signal are subjected to the same detection efficiency. Supported by these observations we will assume in the following analysis that conversion from cavity to output mode is lossless.

\subsubsection*{Measurement of the Q-function}

To assess the efficiency with which we can detect launched signals, we compute the phase space probability distributions of the vacuum and of a small coherent state. This allows us to obtain the correctly scaled Q-functions of traveling states. Under the assumption that states are converted to the output without loss, we can directly obtain the detection efficiency.

\paragraph*{Phase space probability distribution}
In our experiment we measure in-phase and quadrature $\op{I}(t), \op{Q}(t)$ of the field simultaneously with a phase-preserving amplifier. Integrating this complex signal in time yields measurement records $S = I + \ii Q$, which sample a probability distribution in phase space; scaled correctly (see below), this is the Husimi Q-function of the field incident on the detector \citep{Leonhardt1997}. We perform this integration shot-by-shot with an envelope function that cancels detuning in the demodulation and optimizes the signal-to-noise ratio,
\begin{equation}
    S = \int \dif{t} f(t) \exp(-\ii \omega t) S(t),
\end{equation}
where $S(t)$ is the measured signal, $\omega$ is the remaining detuning after demodulation, and $f(t)$ is chosen to be the envelope of the amplitude of a small coherent state that serves as a reference. In the case $g \ll \kappa_\text{out}$,
\begin{equation}
    f(t) = \expe^{-2g^2t/\kappa_\text{out}} - \expe^{-\kappa_\text{out} t/2}.
\end{equation}

\paragraph*{Influence of loss}
We model the inefficiency of our detection chain as a fictional beam splitter placed between the system and an ideal heterodyne detector (i.e., directly after the output resonator) \citep{Leonhardt1997}. If the detection efficiency is $\etadet$, then this beam splitter transmits a fraction $\etadet$ of the outgoing photons to the detector. The density matrix of the output field is transformed as
\begin{equation}
    \tilde{\rho}_{b_\text{out}} = \text{Tr}_{e} \left[ U_\text{bs}
            \left( \rho_{b_\text{out}} \otimes \ket{0}\bra{0}_e \right)
        U^\dagger_\text{bs} \right],
\end{equation}
where $\mode{e}$ is the ``environment'' mode (or vacuum port of the beam splitter), $\text{Tr}_{e}[\cdot]$ is the trace over that mode, and the beam splitter operation is given by
\begin{equation}
    U_\text{bs} = \exp \left( \arccos(\etadet^{1/2})
        (\mode{b}^\dagger_\text{out} \mode{e} - \mode{b}_\text{out} \mode{e}^\dagger)
        \right).
\end{equation}
The Q-function measured in the experiment is that of $\tilde{\rho}_{b_\text{out}}$, and is related to that of $\rho_{b_\text{out}}$ by a shrinking of the I and Q axes by a factor $\etadet^{1/2}$ and a smoothing with a Gaussian filter. In this picture, the Q-function of the vacuum stays unmodified (a Gaussian with standard deviation 1, located at the origin); that of a coherent state $\ket{\alpha=1}$ remains a Gaussian with unmodified width (standard deviation equal to 1), but moved closer towards the origin, $\alpha \mapsto \eta^{1/2}\alpha$.

\paragraph*{Calibration of IQ axes}
We use these observations to scale the raw data correctly. The raw signal (a complex voltage) is proportional to the desired measurement records, $V(t) \propto S(t)$, where the prefactor depends on the total gain/attenuation in the detection chain, and the integration envelope chosen. We determine this prefactor by measuring the vacuum: we integrate the signal in time and bin the results to obtain a probability distribution in phase space. We find a Gaussian centered at the origin, and by setting its width equal to 1 we obtain a scaling factor that we apply uniformly to all states to obtain their correctly scaled Q-functions. The Q-function for the vacuum is shown in Fig.~3c in the main text.

\subsubsection*{Detection efficiency}

We can evaluate the detection efficiency $\eta_\text{det}$ from the Q-function of a known reference state. Obtaining the correctly scaled Q-function of a coherent state with known amplitude (we typically choose $\alpha_0=1$) we find a Gaussian of width 1. By determining the location $\alpha$ in phase space via a 2D Gaussian fit we obtain directly the detection efficiency, $\etadet = (|\alpha|/|\alpha_0|)^2$. The measured Q-function for a coherent state with $\alpha_0 = 1$ is shown in Fig.~3c of the main text. Because of drifts we measure the detection efficiency interleaved with the release of each state. We find, on average, $\etadet = 0.43 \pm 0.04$. Knowledge of this value allows us to quantitatively analyze the preservation of quantum states during the state release.

\subsection{Analysis of traveling quantum states}

\begin{figure*}[tp]
    \center
    \includegraphics{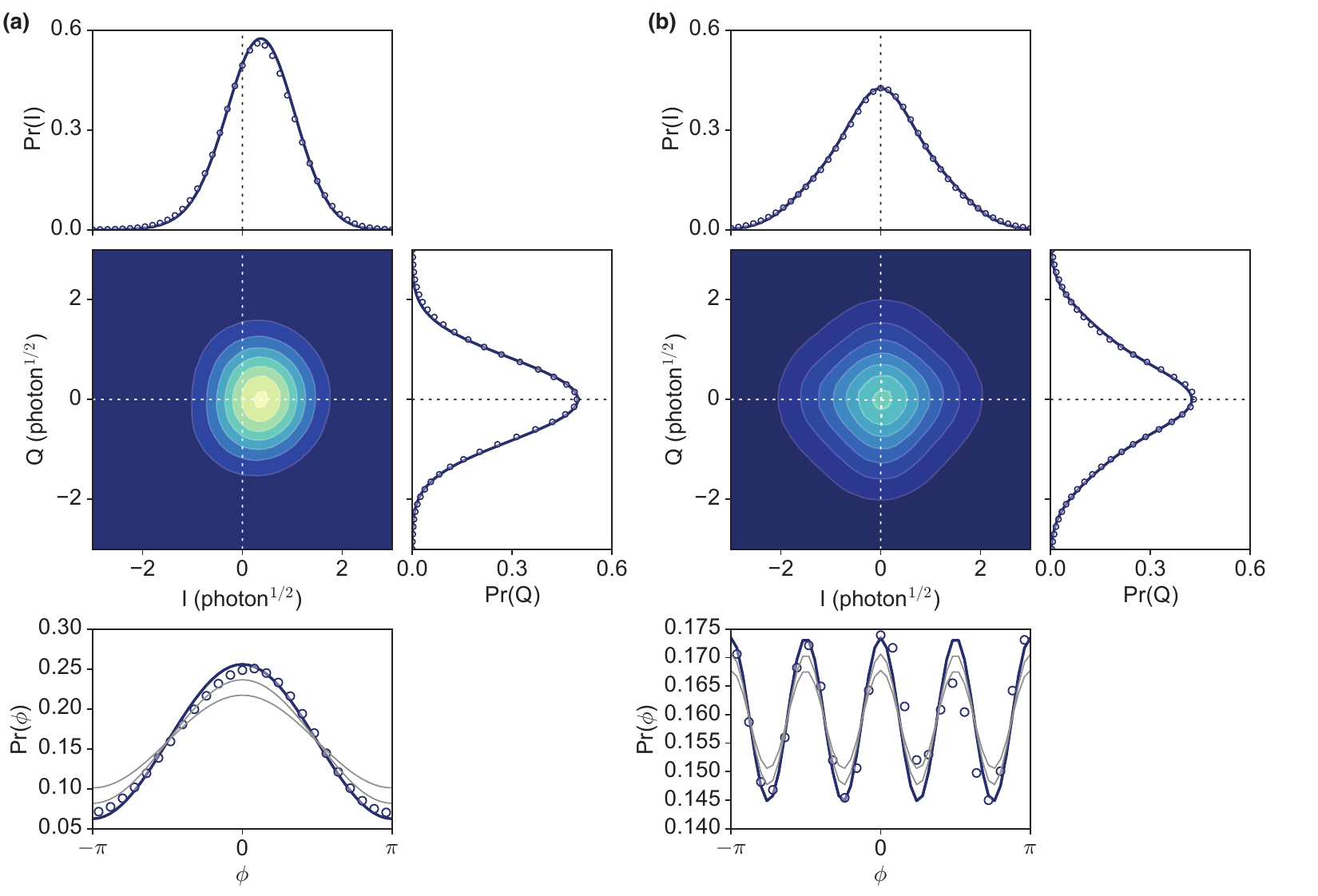}
    \caption{\label{fig:0plusn-fidelity}
        Analysis of the traveling Fock state superpositions. {\bf (a)} $\ket{0}+\ket{1}$ and {\bf (b)} $\ket{0}+\ket{4}$. We show the measured Q-functions as well as the integrated marginals (top and right). Solid lines are the ideal case for the perfect state, only assuming finite detection efficiency (here, $\etadet = 0.47\pm 0.01$). On the bottom, radial integral to obtain the angular probability density function. Dark solid lines: ideal case. As guide to the eye, the light gray lines indicate the expected reduction in contrast if the fidelity was 0.9 and 0.8, respectively (computed by reducing the coherences in the ideal density matrix). Small asymmetries in $\text{Pr}(\phi)$, such as the modulation at $\phi\approx\pi/2$ in the $\ket{0}+\ket{4}$ case, can be explained by state preparation errors.
    }
\end{figure*}

\paragraph*{Preservation of state information in the Q-function}
We launch states with a certain symmetry in phase space and compare the preservation of that symmetry in the Q-function to the ideal case, where only detection efficiency is taken into account. For instance, states of the form $\ket{0}+\ket{n}$ have an $n$-fold symmetry in angular the direction. Importantly, decoherence from photon loss or dephasing beyond the detection efficiency will result in a reduction of contrast. A direct measure of this contrast is to integrate the Q-function radially to obtain the probability density function as a function of polar angle, $\text{Pr}(\phi)$. As a representative example, we show in detail the Q-functions of $\ket{0}+\ket{1}$ and $\ket{0}+\ket{4}$ in Fig.~\ref{fig:0plusn-fidelity}. Based on this type of analysis, we can estimate fidelities to the ideal states exceeding 90\%. The precision of this estimate is limited by our heterodyne detection. We note that this number also depends to $\pm$ 10\% on our assumption that loss is mainly part of the detector as described above.

\paragraph*{Tomography}
It has been shown previously that it is possible to reconstruct the density matrix of a propagating state based on the Q-function, even in the presence of substantial loss \cite{Eichler2012}. We find, however, that the maximum-likelihood methods used to perform this reconstruction introduce large uncertainties in our case. The large Hilbert spaces required for the numerical procedures, together with the available sample sizes, do not allow us to extract information with good precision. We have therefore limited our analysis to the demonstration that essential features in the data are preserved as predicted by theory. The methods shown in this experiment can also be used to capture traveling states (see below); we anticipate that in future experiments highly efficient measurements on traveling fields will be enabled by capturing the signal in a cavity memory and then performing tomography on the cavity state. This will obviate the need for a reconstruction based on the Q-function.

\subsection{Entanglement generation by partial conversion}

\paragraph*{Interaction between cavity and traveling mode}
The interaction between the cavity and the transmission line can be described by an effective beam splitter interaction, where the conversion strength and time determine the coupling $\theta$ between the modes, see Eq.~(\ref{eq:bs-a-bout-theta}). In particular, releasing half the energy stored in the cavity initially corresponds to a `50:50 beam splitter' with $\theta = \pi/2$. The pumping time required to realize this situation can be calibrated precisely with the measurement shown in Fig.~\ref{fig:fock-evolution}.

\paragraph*{Correlation measurements}
In each encoding we perform the following experimental sequence. We first release half the energy of an input state, and record the emitted field during the conversion. Immediately after switching off the pumps, we map the cavity state we wish to probe onto the first excited state of the transmon; we then perform a single-shot, high-fidelity measurement of the transmon state. This mapping and readout thus allow us measure the cavity in any basis of choice. The typical combined mapping and readout fidelity is $\agt 0.95$; we do not correct for finite readout fidelity in any of the data.

\subsubsection*{Half-conversion of a single photon Fock state}

\paragraph*{Measurement protocol}
We start in the state $\ket{1}\ket{0}$; the first ket denotes the state of the cavity, and the second that of the traveling mode. Half-conversion with the operation (\ref{eq:bs-a-bout}) maps this state onto the Bell state
\begin{equation}
    \label{eq:Fock1-bell-state}
    \frac{1}{\sqrt{2}}(\ket{1}\ket{0} + \ket{0}\ket{1}) = \frac{1}{\sqrt{2}} (\ket{+}\ket{+} - \ket{-}\ket{-}),
\end{equation}
where we have defined $\ket{\pm} = (\ket{0}\pm\ket{1})/\sqrt{2}$. To measure in the number basis, we apply a transmon $\pi$-pulse that is selective on either $\ket{0}$ or $\ket{1}$ and then measure in a single shot if the transmon is in the excited state; due to the asymmetry of the readout errors we discard the outcomes in which the transmon is found in the ground state ($\sim 50\,\%$ of the cases). To measure in the $\ket{+}, \ket{-}$ basis, we first map $\ket{+}, \ket{-}$ onto $\ket{0}, \ket{1}$ by an optimized control pulse \cite{Heeres2016}, followed by a measurement in the number basis.

\begin{figure}[tp]
    \center
    \includegraphics{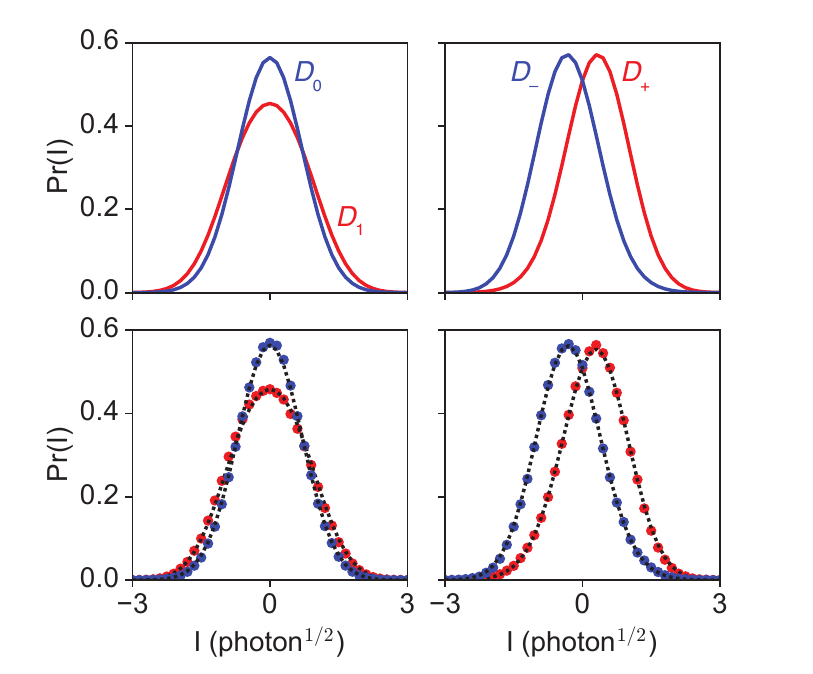}
    \caption{\label{fig:entanglement-01-fit}
        Estimating the fidelity of the entangled state from half-releasing a single photon. Upper panels: Ideal marginals, assuming only the detection efficiency of $0.40\pm0.01$ we have recorded during the half-release experiment. Lower panels: Marginal data conditioned on the cavity outcomes (same as shown in Fig.~4b of the main text). Dashed black lines are fits to the model described in the text. Data have not been corrected for initialization or the conditioning cavity measurements.
    }
\end{figure}

\paragraph*{Lower bound on the Bell state fidelity}
Using a simple model of the detector we can place a lower bound on the fidelity of the entangled state. For a two-qubit Bell state of the form (\ref{eq:Fock1-bell-state}) a strict lower bound on the entangled state fidelity is given by \cite{Sackett2000}
\begin{equation}
    F \geq \frac{1}{2}\left(
        \rho_{22} + \rho_{33} + \tilde\rho_{11} + \tilde\rho_{44} -
            \tilde\rho_{22} - \tilde\rho_{33} - 2\sqrt{\rho_{11}\rho_{44}}
    \right),
\end{equation}
where $\rho$ is the density matrix of the joint state in the number basis, and $\tilde\rho$ is the density matrix in the $\ket{+}, \ket{-}$ basis. This lower bound can be evaluated with the data shown in Fig.~4b in the main text. The diagonal density matrix elements in the above expression can be found from the probabilities with which we find the cavity and propagating modes in a particular state, i.e.,
\begin{equation}
\begin{aligned}
    F \geq \frac{1}{2} \Big(&
        P_a(0)P_b(1|0_a) + P_a(1)P_b(0|1_a)\\
        & - 2\sqrt{P_a(0)P_b(0|0_a) P_a(1)P_b(1|1_a)} \\
        & + P_a(+)P_b(+|+_a) + P_a(-)P_b(-|-_a) \\
        & - P_a(-)P_b(+|-_a) - P_a(+)P_b(-|+_a)
    \Big).
\end{aligned}
\end{equation}
Here, $P_a(i)$ is the probability to find the cavity in state $\ket{i}$ in a measurement in the $\ket{i}, \ket{\bar{i}}$ basis, and $P_b(j|i_a)$ is the conditional probability to find the propagating state in $\ket{j}$ after having found the cavity in $\ket{i}$.

\paragraph*{Extraction of the lower bound from the field data}
The probabilities $P_a(\cdot)$ follow directly from the statistics of the cavity measurements, which leaves us with the task to find the conditional probabilities for the traveling state. We compute these by fitting the measured marginals $\text{Pr}(I)$ (Fig.~\ref{fig:entanglement-01-fit}). With known detection efficiency we can compute the ideal marginals, $D_{0,1}$ in the number basis, and $D_{+,-}$ in the rotated basis. Assuming our detector is linear, we can then fit the conditioned field data in the $\ket{i}, \ket{\bar{i}}$ basis to $\alpha D_{i} + (1-\alpha) D_{\bar{i}}$, with $0 \leq \alpha \leq 1$. The conditional probabilities are then determined by the single fit parameter, $P_b(i|i_a) = \alpha$. Assuming again that the conversion is lossless and we can reliably determine the detection efficency, we find a lower bound for the entangled state fidelity of $F \geq 0.91\pm0.02$, uncorrected for any inefficiency in the preparation or measurement of the cavity state.

\subsubsection*{Half-conversion of a two-photon Fock state}

\begin{figure}[tp]
    \center
    \includegraphics{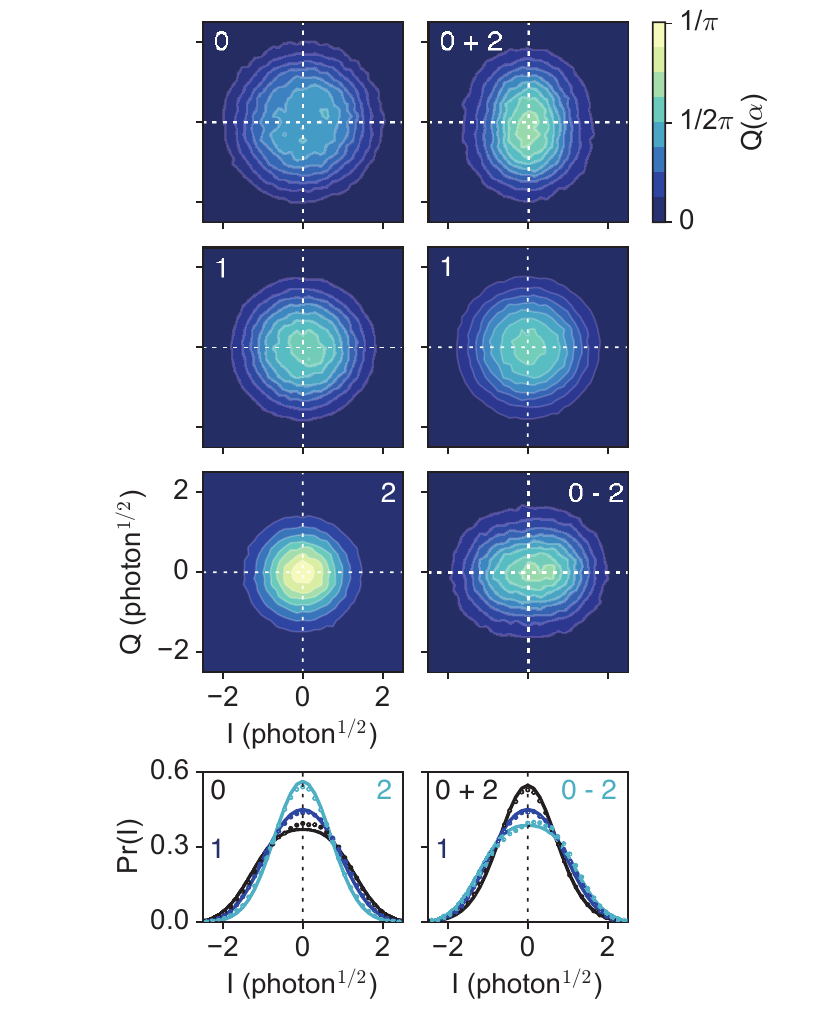}
    \caption{\label{fig:half-pitch-fock2}
        Conditioned Q-functions after half-release of $\ket{2}$. Left column: Q-functions conditioned on finding in the cavity in either $\ket{0}$, $\ket{1}$, or $\ket{2}$; for these case we expect the field to be in either $\ket{2}$, $\ket{1}$, or $\ket{0}$, respectively. Right column: conditioned on finding the cavity in $\ket{+}_2$, $\ket{1}$, or $\ket{-}_a$. Solid lines in the marginals: Ideal case, only detection efficiency for the traveling field taken into account.
    }
\end{figure}

The entangled states arising from half-releasing Fock states larger than $\ket{1}$ are not maximally entangled states in the number basis, but still display non-classical correlations. We illustrate this by half-releasing the Fock-state $\ket{2}$, which is mapped onto
\begin{equation}
\begin{aligned}
    & \frac{1}{\sqrt 2}\ket{1}\ket{1} + \frac{1}{2} (\ket{2}\ket{0} + \ket{0}\ket{2}) \\
    =\, &\frac{1}{\sqrt 2}\ket{1}\ket{1} + \frac{1}{2} (\ket{+}_2\ket{+}_2 - \ket{-}_2\ket{-}_2),
\end{aligned}
\end{equation}
where now $\ket{\pm}_2 = (\ket{0}\pm\ket{2})/\sqrt{2}$. To reveal correlations we can thus probe the states $\ket{0}, \ket{1}, \ket{2}$ (by number-selective $\pi$-pulses and transmon measurement), and $\ket{+}_2, \ket{1}, \ket{-}_2$ (mapping onto $\ket{0}, \ket{1}, \ket{2}$ by a single combined transmon/cavity pulse, followed by number-selective $\pi$-pulses) (Fig.~\ref{fig:half-pitch-fock2}).

\subsubsection*{Half-conversion of a cat state}

\begin{figure}[tp]
    \center
    \includegraphics{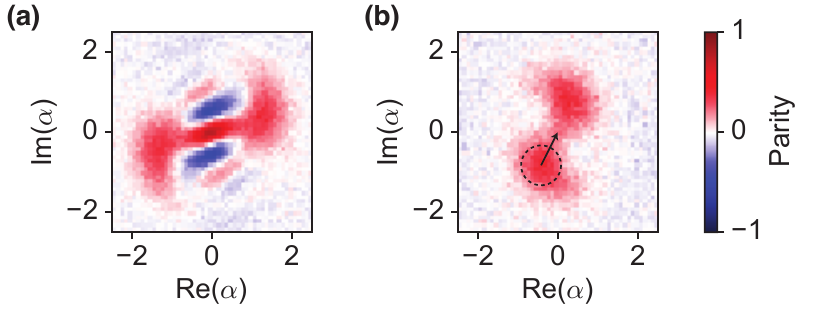}
    \caption{\label{fig:cat-kerr}
        Kerr effect during half-release of a cat-state.
        {\bf (a)} Cavity state after creating $\catket{+}{\sqrt{2}}$ followed by 3\,$\mu$s of delay. While the state maintains its purity, the Kerr effect leads to a rotation and a `smearing' of the populations.
        {\bf (b)} Cavity state after half-release of $\catket{+}{\sqrt{2}}$. The state now appears fully mixed due to the entanglement with the traveling mode; the additional rotation with respect to the case of {\bf (a)} comes from the Stark shifts induced by the pumps. Mapping onto the transmon is achieved by displacing one of the `blobs' to the vacuum followed by mapping the vacuum onto the transmon excited state. This is indicated by the circle and arrow; the radius of the circle is 1/2, which is the standard deviation of a coherent state in the Wigner function. It can readily be seen that the smearing leads to a small imperfection in the mapping because the blob will not be confined to the vacuum after the displacement.
    }
\end{figure}

\paragraph*{Protocol}
Because a coherent state $\ket{\sqrt 2 \alpha} \ket{0}$ is mapped onto $\ket{\alpha}\ket{\alpha}$, we can readily see that the cat state $\catket{+}{\sqrt{2}}\ket{0}$ results in a two-mode entangled cat state
\begin{equation}
\begin{aligned}
    & \mathcal{N}_\alpha(\ket{1}_\alpha\ket{1}_\alpha + \ket{-1}_\alpha\ket{-1}_\alpha) \\
    =\, & \mathcal{N}_\text{c} (\catket{+}{1}\catket{+}{1} + \catket{-}{1}\catket{-}{1}).
\end{aligned}
\end{equation}
Here, $\mathcal{N}_{\alpha,\text{c}}$ are normalization factors, and we use the notation $\ket{\beta}_\alpha$ to denote a coherent state with complex amplitude $\beta$. We expect correlations in the coherent state phase and photon number parity bases, because the cat states $\catket{\pm}{1}$ are eigenstates of even ($+$) and odd ($-$) parity, respectively. To measure in the coherent state basis, we displace the cavity state by $\alpha = 1$, followed by a transmon $\pi$-pulse that is selective on the vacuum; $\ket{-1}_\alpha$ is thus mapped onto the the transmon excited state, while $\ket{1}_\alpha$ is mapped onto the ground state. Subsequent transmon measurement thus gives the result in the $\ket{\alpha}, \ket{\bar{\alpha}}$ basis. To measure parity we apply a series of $\pi$-pulses that are selective on the even photon numbers on the transmon; this maps even parity onto the excited state, and odd parity onto the ground state.

We note that the states $\ket{\alpha=\pm1}$ have a non-zero overlap ($\sim 4\,\%$). This means that the coherent state basis is not perfectly orthogonal for the cat state $\catket{\pm}{1}$. This overlap explains why the conditioned Q-functions after measuring the cavity in the $\ket{\alpha}, \ket{\bar\alpha}$ basis are slightly asymmetric. This also prevents us from using the simple method we have employed to bound the entangled state fidelity in the single Fock-state case.

\begin{figure}[tp]
    \center
    \includegraphics{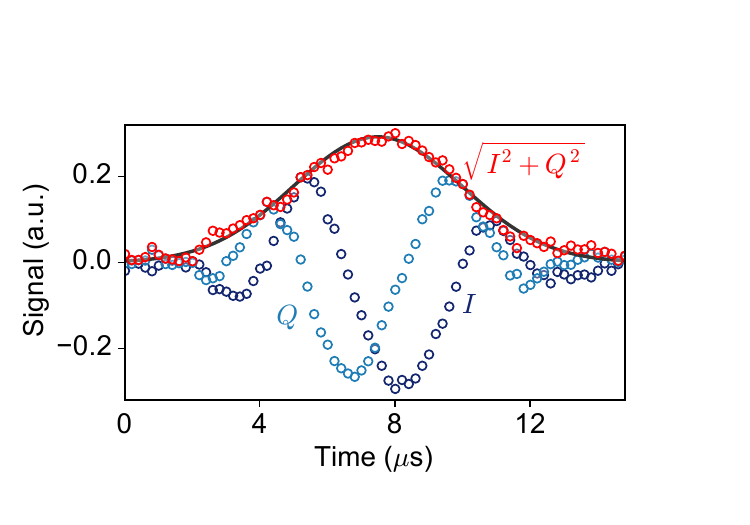}
    \caption{\label{fig:shaped}
        Example of a Gaussian shaped wavepacket. Given the system Hamiltonian, we numerically solve the equations of motion which dictate the shape of $\langle \modebout (t)\rangle$ in terms of the drives $\xi_{1,2}(t)$, which are amplitude and phase modulated. Solid line is the envelope of the desired wavepacket. Red points are the envelope of the measured output field, while dark and light blue points are the in-phase ($I$) and  quadrature ($Q$), respectively. The desired and measured signals are scaled to have the same integral.
    }
\end{figure}

\paragraph*{Cavity evolution due to the Kerr-effect}
The single-photon Kerr-effect in the cavity \cite{Haroche2006,Kirchmair2013} is described by the Hamiltonian
\begin{equation}
    H_\text{Kerr} = \frac{\chi_{aa}}{2} \modea^{\dagger 2} \modea^2.
\end{equation}
This nonlinearity causes a unitary evolution in which each photon number acquires phase at a rate $n\chi_{aa}$. On long time scales the state undergoes cyclic evolution at revival times $2\pi/\chi_{aa}$. On shorter times, the evolution leads to `smearing' of the phase information, with a phase collapse time that depends on the mean number of photons, $\bar{n}$, in the cavity, $T_{\text{collapse}} = \pi/(2 \sqrt{\bar{n}} \chi_{aa})$. In our system, $\chi_{aa} = 2\pi \times 22\,\text{kHz}$, and thus $T_{\text{collapse}} \approx \bar{n}^{-1/2} \times 11.4 \,\mu\text{s}$.

\paragraph*{Reduction of measurement contrast due to Kerr evolution}
The total time in between creating the cat state and mapping the coherent state onto the transmon for measurement in the $\ket{\pm\alpha}$ basis is 3\,$\mu$s. The Kerr evolution during this time leads to slight smearing of the populations at $\alpha = \pm 1$, resulting in an imperfection of the mapping onto the transmon (Fig.~\ref{fig:cat-kerr}a,b). This imperfection is asymmetric: the smearing leads to a reduced probability of the transmon being flipped conditional on the vacuum, but a finding the transmon excited is more likely to give the correct outcome than finding it in the ground state. This imperfection explains the deviation of the data from the ideal distribution shown in Fig.~4c of the main text. The Kerr effect does not alter photon number parity, and the correlations in that basis are closer to the ideal case. Undesired Kerr evolution can be suppressed in future experiments by reducing the magnitude of the Kerr effect and by increasing the bandwidth of the conversion process.

\section{Shaping the outgoing wavepacket}
Amplitude and phase control of the drives applied to the system in principle allow for control over the shape of the outgoing wavepacket; this shaping is a necessary ingredient for deterministic state transfer protocols \cite{Cirac1997}. The control problem in our system is defined by the system Hamiltonian (\ref{eq:stark_hamiltonian}) and the resulting equations of motion (\ref{eq:equationsofmotion}). Amplitude control of the drives $\xi_{1,2}$ allow for temporal shaping of the conversion term $g(t)$, while phase control allows for dynamic frequency tuning. This is necessary because stark shifts are induced on all modes by the drive tones. To keep the mixing process resonant, the drives must be frequency compensated. We numerically solve the equations of motion for the drives $\xi_{1,2}$ to generate a specified output wavepacket $\modebout (t)$. Figure~\ref{fig:shaped} shows an example of a shaped wavepacket, which demonstrates the ability to generate wavepackets of a specified shape in the presence of dynamic frequency shifts. Because emission is the time-reverse of capture, this control directly translates to the ability to capture shaped wavepackets.





%

\end{document}